\documentstyle[preprint,aps,epsf]{revtex}
\def\lessim{\mathrel {\vcenter {\baselineskip 0pt \kern 0pt
\hbox{$<$} \kern 0pt \hbox{$\sim$} }}}
\def\gessim{\mathrel {\vcenter {\baselineskip 0pt \kern 0pt
\hbox{$>$} \kern 0pt \hbox{$\sim$} }}}
\def \rightdownarrow
 {\kern.4em {\raise1.75ex\hbox{$\Bigl |$}$\kern-0.27em{\longrightarrow}$}}
\def \mrightdownarrow
 {\kern.4em {\raise1.19ex \hbox{$|$} \kern-0.23em{
                                                \longrightarrow}}}
\def\Pt{$P_T$}
\def\GeVc{GeV$\!/c$}
\def\GeVcc{GeV$\!/c^2$}
\def\Bu{$B^+$}
\def\Bd{$B^0$}
\def\Bs{$B_s^0$}
\def\mJpsi{J/\psi}
\def\Jpsi{$\mJpsi$}
\def\psiprime{$\psi(2S)$}
\def\mpsiprime{\psi(2S)}
\def\pbarp{$p\bar{p}$}

\def\mKstarzero{K^{\ast}(892)^0}
\def\Kstarzero{$\mKstarzero$}
\def\mBR{{\cal B}}
\def\BR{$\mBR$}

\def\invpb{${\rm pb^{-1}}$}
\def\invcms{$\rm cm^{-2}s^{-1}$}

\def\etal{{\it et al.}}
\def\r#1{\ignorespaces $^{#1}$}
\begin{document}
\tightenlines
\title{
\begin{flushright}
FERMILAB-Pub-98/091-E \\
Phys. Rev. D {\bf 58}, 072001 (1998) \\
%
%
%
\end{flushright}
Observation of $B^+\rightarrow \mpsiprime \, K^+$\ and $B^0\rightarrow
\mpsiprime \, K^\ast(892)^0$\ decays and measurements of $B$-meson
branching fractions into \Jpsi\ and \psiprime\ final states
}
\author{
\font\eightit=cmti8
\hfilneg
\begin{sloppypar}
\noindent
F.~Abe,\r {17} H.~Akimoto,\r {39}
A.~Akopian,\r {31} M.~G.~Albrow,\r 7 A.~Amadon,\r 5 S.~R.~Amendolia,\r {27} 
D.~Amidei,\r {20} J.~Antos,\r {33} S.~Aota,\r {37}
G.~Apollinari,\r {31}
% ...
\end{sloppypar}
\vskip .026in
\begin{center}
(CDF Collaboration)
\end{center}
\newpage
\vskip .026in
\begin{center}
\r 1  {\eightit Argonne National Laboratory, Argonne, Illinois 60439} \\
\r 2  {\eightit Istituto Nazionale di Fisica Nucleare, University of Bologna,
I-40127 Bologna, Italy} \\
\r 3  {\eightit Brandeis University, Waltham, Massachusetts 02254} \\
\r 4  {\eightit University of California at Los Angeles, Los 
Angeles, California  90024} \\  
\r 5  {\eightit University of Chicago, Chicago, Illinois 60637} \\
\r 6  {\eightit Duke University, Durham, North Carolina  27708} \\
\r 7  {\eightit Fermi National Accelerator Laboratory, Batavia, Illinois 
60510} \\
\r 8  {\eightit University of Florida, Gainesville, Florida  32611} \\
\r 9  {\eightit Laboratori Nazionali di Frascati, Istituto Nazionale di Fisica
               Nucleare, I-00044 Frascati, Italy} \\
\r {10} {\eightit University of Geneva, CH-1211 Geneva 4, Switzerland} \\
\r {11} {\eightit Harvard University, Cambridge, Massachusetts 02138} \\
\r {12} {\eightit Hiroshima University, Higashi-Hiroshima 724, Japan} \\
\r {13} {\eightit University of Illinois, Urbana, Illinois 61801} \\
\r {14} {\eightit Institute of Particle Physics, McGill University, Montreal 
H3A 2T8, and University of Toronto,\\ Toronto M5S 1A7, Canada} \\
\r {15} {\eightit The Johns Hopkins University, Baltimore, Maryland 21218} \\
\r {16} {\eightit Institut f\"{u}r Experimentelle Kernphysik, 
Universit\"{a}t Karlsruhe, 76128 Karlsruhe, Germany} \\
\r {17} {\eightit National Laboratory for High Energy Physics (KEK), Tsukuba, 
Ibaraki 305, Japan} \\
\r {18} {\eightit Ernest Orlando Lawrence Berkeley National Laboratory, 
Berkeley, California 94720} \\
\r {19} {\eightit Massachusetts Institute of Technology, Cambridge,
Massachusetts  02139} \\   
\r {20} {\eightit University of Michigan, Ann Arbor, Michigan 48109} \\
\r {21} {\eightit Michigan State University, East Lansing, Michigan  48824} \\
\r {22} {\eightit University of New Mexico, Albuquerque, New Mexico 87131} \\
\r {23} {\eightit The Ohio State University, Columbus, Ohio 43210} \\
\r {24} {\eightit Osaka City University, Osaka 588, Japan} \\
\r {25} {\eightit Universita di Padova, Istituto Nazionale di Fisica 
          Nucleare, Sezione di Padova, I-35131 Padova, Italy} \\
\r {26} {\eightit University of Pennsylvania, Philadelphia, 
        Pennsylvania 19104} \\   
\r {27} {\eightit Istituto Nazionale di Fisica Nucleare, University and Scuola
               Normale Superiore of Pisa, I-56100 Pisa, Italy} \\
\r {28} {\eightit University of Pittsburgh, Pittsburgh, Pennsylvania 15260} \\
\r {29} {\eightit Purdue University, West Lafayette, Indiana 47907} \\
\r {30} {\eightit University of Rochester, Rochester, New York 14627} \\
\r {31} {\eightit Rockefeller University, New York, New York 10021} \\
\r {32} {\eightit Rutgers University, Piscataway, New Jersey 08855} \\
\r {33} {\eightit Academia Sinica, Taipei, Taiwan 11530, Republic of China} \\
\r {34} {\eightit Texas A\&M University, College Station, Texas 77843} \\
\r {35} {\eightit Texas Tech University, Lubbock, Texas 79409} \\
\r {36} {\eightit Istituto Nazionale di Fisica Nucleare, University of
Trieste/Udine, Italy} \\
\r {37} {\eightit University of Tsukuba, Tsukuba, Ibaraki 315, Japan} \\
\r {38} {\eightit Tufts University, Medford, Massachusetts 02155} \\
\r {39} {\eightit Waseda University, Tokyo 169, Japan} \\
\r {40} {\eightit University of Wisconsin, Madison, Wisconsin 53706} \\
\r {41} {\eightit Yale University, New Haven, Connecticut 06520} \\
\end{center}
}
\draft
\address{}
\date{17 March 1998}
\maketitle
%
%
\newpage
\begin{abstract}
We report the observations of the decays $B^+ \rightarrow \mpsiprime
\, K^+$\ and $B^0 \rightarrow \mpsiprime \, K^\ast(892)^0$\ in \pbarp\
collisions at a center-of-mass energy of 1.8 TeV using a 110~\invpb\
data sample recorded by the Collider Detector at Fermilab.  We also
reconstruct the decays $B^+ \rightarrow \mJpsi \, K^+$\ and $B^0
\rightarrow \mJpsi \, K^\ast(892)^0$ and measure the six ratios of
branching fractions of these four decays.  The relative
branching-fraction results are shown to be consistent with
phenomenological factorization calculations of hadronic $B$-meson
decays.  We use the world-average branching fraction ${\cal B}(B^+\to
J/\psi\,K^+)$ to derive $\mBR(B^+\to \mpsiprime\, K^+)=
(0.56\pm0.08\pm0.10)\times 10^{-3}$, $\mBR(B^0 \to
\psi(2S)\,K^\ast(892)^0)= (0.92\pm0.20\pm0.16)\times 10^{-3}$, and
$\mBR(B^0\rightarrow \mJpsi\, K^\ast(892)^0)=
(1.78\pm0.14\pm0.29)\times10^{-3}$, where the first and second
uncertainties are statistical and systematic, respectively.
\end{abstract}

%
%
\pacs{PACS Numbers: 13.25.Hw, 13.87.Fh, 14.40.Nd}

%
\section{Introduction}

Studies of the decays of bound states of bottom quarks and light
antiquarks have proven to be one of the most effective ways to explore
the decay dynamics of heavy quark systems.  The branching fractions
(\BR) of the decays of the two lowest-lying bound states, the \Bu\ and
\Bd\ mesons, depend on a blend of effects due to the weak and strong
interactions.  The measurements of semileptonic decays of $B$ mesons,
where a charged lepton and its corresponding neutrino are produced,
have proven to be useful in the development of theoretical models that
relate the semileptonic branching fractions to each other~\cite{ref:
BSW semileptonic}.  In a similar way, the measurements of the fully
hadronic decays (modes where the $B$-meson decay daughters are
hadrons) have also been shown to provide tests of the theory of heavy
quark decay~\cite{ref: BSW hadronic,ref: Browder etal}.

In this paper, we report the observations of the decays
$B^+\rightarrow \mpsiprime \, K^+$\ and $B^0\rightarrow \mpsiprime \,
K^\ast(892)^0$\ in \pbarp\ collisions at a center-of-mass energy of
1.8~TeV.  We also observe the decays $B^+\rightarrow \mJpsi \, K^+$\
and $B^0\rightarrow \mJpsi \, K^\ast(892)^0$\ and use our data to
measure the ratios of branching fractions of the \Bu\ and \Bd\ mesons
to these four final states.  We compare our measurements with
theoretical branching-fraction ratio predictions.  Throughout this
paper, references to a specific decay mode imply the charge conjugate
mode as well.

\begin{figure}
\begin{center}
\leavevmode
\hbox{%
\epsfxsize=3.4in
\epsffile{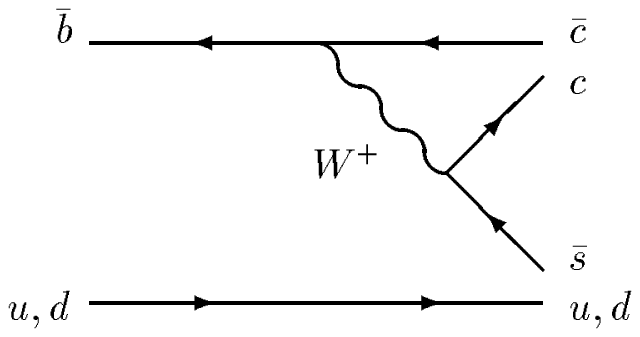}}
\end{center}
\caption
{Diagram of the color-suppressed internal-$W$ emission
mechanism for a $B$ meson (here either a $\bar{b}u$ ($B^+$) or
$\bar{b}d$ ($B^0$) state) decaying to charmonium ($J/\psi$
or $\psi(2S)$) and a strange meson ($K^+$ or $K^*(892)^0$).
In this process, the $u$ or $d$ quark is assumed to be a
``spectator'' of the weak interaction.}
\label{fig:bdecay_feyn}
\end{figure}

As illustrated in the diagram in Fig.~\ref{fig:bdecay_feyn}, all four
of the \Jpsi\ and \psiprime\ decay modes are color-suppressed
Cabibbo-favored decays; they can only occur when the $W$ boson's
hadronic decay products, themselves a color singlet, combine with the
charm antiquark from the flavor-changing decay and the light spectator
quark to form color-singlet charmonium and strange mesons,
respectively.  Strong interaction effects, however, are expected to
modify the dynamics of these decays.  The most successful theoretical
treatments of such decays employ the factorization hypothesis, where
the decay of the $B$ meson is described by processes that take place
on different time scales: short-distance hard-gluon exchange and the
weak nonleptonic decay of the $b$~quark, followed by longer-distance
strong interactions between the final-state partons that produce the
two daughter mesons.  The decay amplitude is factorized into a product
of hadronic currents that reduces to the charmonium decay constant and
the matrix element for the $B \to K$ hadronic current, which consists
of several form factors~\cite{ref: BSW semileptonic,ref: BSW
hadronic}.  Measurements of the rates and polarization of these decays
confront the assumptions that underlie the factorization hypothesis in
$B$-meson decays and the calculations involving hadronic form factors.

Exclusive hadronic decays of $B$~mesons are difficult to observe due
to their relatively small branching fractions (typically $10^{-4}$ to
$10^{-2}$) and the small exclusive branching fractions of the
subsequent charm daughter decays.  However, the large production cross
section for bottom quarks in \pbarp\ collisions (in the range of 2-3
$\mu$b for quarks with transverse momentum $P_T>6$~\GeVc\ and rapidity
$|y_b|<1$) has made it possible to identify relatively large samples
of specific decay modes~\cite{ref: CDF B sigma}.  The decay modes that
involve a charmonium daughter meson have proven to be most amenable to
study, as the decay of the charmonium state involving two energetic
muons yields a distinctive signature that can be used to identify the
candidate events.  The Collider Detector at Fermilab (CDF)
Collaboration has published a number of measurements of the properties
of $B$~mesons using final states involving a \Jpsi\ meson, including
production cross sections, masses, lifetimes, polarizations, and
branching fractions~\cite{ref: CDF B sigma,ref: J/psi Bs mass,ref: Run
1 B lifetime,ref: J/psi polarization,ref: Run 1A BR}.

The CDF Collaboration measured the branching fractions of \Bu, \Bd,
and \Bs\ mesons using five different decay modes, all identified by
requiring a $\mJpsi\rightarrow\mu^+\mu^-$\ decay~\cite{ref: Run 1A
BR}.  We have now completed a more extensive
study, incorporating a factor of four more data and focusing on the
final states identified by the presence of a decay of the form
\begin{eqnarray}
B &\rightarrow& \mpsiprime \, K \\
  && \ \mrightdownarrow \,\mu^+\,\mu^- \nonumber \\
  && \ \mrightdownarrow \,\mJpsi \,\pi^+\,\pi^- \nonumber \\
  && \quad \quad \quad \mrightdownarrow \,\mu^+\,\mu^-, \nonumber
\end{eqnarray}
where $B$\ is a \Bu\ (\Bd) meson and $K$\ is a $K^+$\ (\Kstarzero)
meson.  The \Kstarzero\ meson is observed through its decay to the
$K^+\pi^-$\ final state.

We first describe in Secs.~\ref{sect: experimental approach} and
\ref{sect: data collection} the experiment and data-collection
procedures used for this measurement.  In Sec.~\ref{sect: event
selection}, we discuss the event selection procedure and present the
observed rates of the various $B$-meson decays.  The necessary
efficiency corrections to convert the observed decay rates to
branching fractions are discussed in Sec.~\ref{sect: efficiency
corrections} where we also detail the systematic uncertainties
associated with the measurement.  We present our results in
Sec.~\ref{sect: results} and offer our conclusions in Sec.~\ref{sect:
conclusion}.

\section{Experimental Approach}
\label{sect: experimental approach}
In principle, the number of observed events for the decay mode $ B^+
\rightarrow \mpsiprime \,K^+$\ can be decomposed into the form
\begin{eqnarray}
N_{\rm obs}(\mpsiprime \,K^+)
 = 2 \int\!\!{\cal L}dt \cdot \sigma(p\bar{p}\rightarrow \bar{b})
                    \cdot f_u\cdot
                    \mBR(B^+ \rightarrow \mpsiprime \,K^+)\cdot
			\epsilon^{\psi(2S)\,K^+},
\label{eq: cross section}
\end{eqnarray}
and similar forms can be written for the other decays.  Here,
$\int\!\!{\cal L}dt$ is the time-integrated luminosity,
$\sigma(p\bar{p}\rightarrow \bar{b})$\ is the bottom antiquark
production cross section, and $f_u$\ is the probability that the
fragmentation of a bottom antiquark will result in a \Bu\ meson.  In a
similar way, we define $f_d$\ to be the probability of a bottom
antiquark to hadronize and form a \Bd\ meson.  We refer to these
probabilities as fragmentation fractions and include in these
fractions contributions from decays of heavier bottom hadrons into
final states containing a \Bu\ or \Bd\ meson.  The expression
$\mBR(B^+ \rightarrow \mpsiprime \, K^+)$\ represents the branching
fraction for this $B^+$-meson decay mode, and
$\epsilon^{\psi(2S)\,K^+}$\ is the acceptance and efficiency of
detecting the $\mpsiprime \,K^+$\ final state.  The factor of 2
accounts for the possibility of reconstruction of $B^-$ or $B^+$
mesons in each event.

The observation of a certain number of $B$-meson decays in a specific
mode can be converted into a branching-fraction measurement using an
expression similar to Eq.~(\ref{eq: cross section}).  However,
uncertainties in the bottom-quark production cross section~\cite{ref:
CDF B sigma} can be avoided by measuring ratios of branching fractions
between $B$-meson decay modes, a procedure that also results in the
beneficial cancellation of several detection and reconstruction
efficiencies and their associated uncertainties.  For example, a
branching-fraction ratio involving only $J/\psi$ charmonium mesons is
measured as
\begin{equation}
\frac{{\cal B}(B^0\to J/\psi\,K^*(892)^0)}{{\cal B}(B^+\to J/\psi\,K^+)}
=
\frac{N_{\rm obs}(J/\psi\,K^*(892)^0) \cdot f_u \cdot \epsilon^{J/\psi\,K^+}}
     {N_{\rm obs}(J/\psi\,K^+) \cdot f_d \cdot \epsilon^{J/\psi\,K^\ast(892)^0}
     \cdot
     {\cal B}(K^*(892)^0\to K^+\,\pi^-)},
\end{equation}
where $N_{\rm obs}(J/\psi\,K^*(892)^0)$ and $N_{\rm obs}(J/\psi\,K^+)$ denote
the observed event yields and $\epsilon^{J/\psi\,K^*(892)^0}$ and
$\epsilon^{J/\psi\,K^+}$ represent the detector acceptance and
reconstruction efficiencies, which have several common factors, such
as the $J/\psi$ branching fraction and reconstruction efficiency, that
divide out of the ratio.  Ratios of branching fractions are also
beneficial in theoretical studies of these decays since several common
factors divide out of the amplitude expressions~\cite{ref:
gourdin_etal_aleksan_etal,ref: kamal_santra}.

These ratios of branching fractions can be used to estimate absolute
branching fractions using world-average values for the denominator of
the ratios.  This is particularly useful for those ratios that involve
the most precisely known branching fraction, $\mBR(B^+\to J/\psi\,
K^+)$, in the denominator.  We use the world-average value of
$\mBR(B^+\to J/\psi\, K^+)$ to estimate the absolute branching
fractions of the other three decay modes.

\section{Data Collection}
\label{sect: data collection}
\subsection{The CDF Detector}
The Collider Detector at Fermilab is a multi-purpose detector designed
to study 1.8~TeV \pbarp\ collisions produced by the Fermilab Tevatron
collider~\cite{ref: CDF Detector}.  The detector has a coordinate
system with the $z$\ axis along the proton beam direction, the $y$\
axis pointing vertically upwards, and the $x$\ axis pointing
horizontally.  The polar angle $\theta$\ is defined relative to the
$z$\ axis and $\phi$\ is the azimuthal angle.  Pseudorapidity is
defined as $\eta\equiv-\ln[\tan(\theta/2)]$.  The CDF detector
surrounds the interaction region with three charged-particle tracking
detectors immersed in a 1.4~T solenoidal magnetic field.  The tracking
system is contained within a calorimeter system that measures the
energy flow of charged and neutral particles out to $|\eta|<4.2$.
Charged-particle detectors outside the calorimeter are used to
identify muon candidates.

The innermost tracking device is a silicon microstrip detector (SVX)
located in the region between 2.9 and 7.9 cm in radius from the beam
axis.  The SVX is surrounded by a set of time projection chambers
(VTX) that measure charged-particle trajectories to a radius of 22 cm.
An 84 layer drift chamber (CTC) measures the particle trajectories in
the region between 30 and 132 cm in radius from the beam.  This
tracking system has high efficiency for detecting charged particles
with momentum transverse to the beam $P_T > 0.35$~\GeVc\ and
$|\eta|\lessim1.1$.  Together, the CTC and SVX measure
charged-particle transverse momenta with a precision of $\sigma_{P_T}
\simeq [(0.0066 {P_T})^2 + (0.0009 P_T^2)^2]^{1/2}$\ (with \Pt\ in
units of \GeVc).

The muon detection system has four of its layers of planar drift
chambers separated from the interaction point by approximately five
interaction lengths of material.  To reduce the probability of
misidentifying penetrating hadrons as muon candidates in the central
pseudorapidity region $|\eta|<0.7$, four more layers of chambers are
located outside the magnet return yoke (corresponding to a further
three interaction lengths of material at $\theta=90^\circ$).  An
additional set of chambers is located in the pseudorapidity interval
$0.7<|\eta|<1.0$\ to extend the polar acceptance.  The muon system is
capable of detecting muons with $P_T \gessim 1.4$\ \GeVc\ in a
pseudorapidity interval $|\eta| < 1.0$.  A three-level trigger system
is used to select candidate collisions for subsequent study.  These
and other elements of the CDF detector are described in more detail
elsewhere~\cite{ref: CDF Detector}.

\subsection{The Data Set}
The data were collected in two running periods, the first extending
for nine months starting in August 1992, and the second extending for
18 months starting in January 1994.  Several modifications were made
to the CDF detector during the hiatus between these two running
periods.  The most significant of these were the replacement of the
silicon microstrip detector, the commissioning of a trigger system
with greater selectivity, and improvements to the data acquisition
system.  The most notable difference in running conditions resulted
from a rise in the average instantaneous luminosity of the Tevatron
accelerator.  The mean instantaneous luminosities during the two
periods were $3.5\times 10^{30}$~\invcms\ and
$8.0\times10^{30}$~\invcms, respectively, and the peak instantaneous
luminosity exceeded $2.0\times10^{31}$~\invcms.  The time-integrated
luminosity of the data sample for the two running periods is
$\sim$20~\invpb\ and $\sim$89~\invpb, respectively.

Despite the differences in detector configuration during the two
running periods, we were able to treat the two sets of data as a
single sample with a total time-integrated luminosity of
$\sim$109~\invpb.  This was achieved through the use of nearly
identical event reconstruction techniques and consistent calibration
procedures for data collected during the two running periods.

\section{Event Selection}
\label{sect: event selection}

\subsection{The $J/\psi$\ and \psiprime\ Trigger Requirements}
A common feature of the four $B$-meson decay modes studied here 
is the presence of
a $\mu^+\mu^-$\ candidate consistent with that arising from the decay of a
charmonium (\Jpsi\ or \psiprime) state.

We used a three-level trigger system to identify collisions producing
two or more muon candidates.  The first-level trigger required that
two muon candidates be observed in the muon system.  The first-level
trigger track efficiency in the muon system rose from $\sim$40\% at
$P_T=1.5$~\GeVc\ to $\sim$93\% for muons with $P_T>3.0$~\GeVc.  The
second-level trigger required the detection of at least one charged
track in the CTC using the central fast track processor
(CFT)~\cite{ref: CFT NIM}, which performed a partial reconstruction of
all charged tracks with a transverse momentum exceeding $\sim$2~\GeVc.
The CFT track was required to match within $\sim$8$^\circ$\ in $\phi$\ of
the muon candidate.  The CFT efficiency rose from $\sim$40\%\ at a
muon $P_T\sim2$~\GeVc\ to $\sim$94\%\ for $P_T\gessim 3$~\GeVc.  The
third-level trigger required that two reconstructed CTC tracks be
matched with two tracks in the muon chambers and that the invariant
mass of the dimuon pair be between 2.7 and 4.1~\GeVcc.  The efficiency
of the third-level trigger requirement was $(97\pm2)$\%\ for \Jpsi\
decays passing the first and second-level triggers.  Deviations from
these nominal trigger efficiencies were observed to occur during data
acquisition and were taken into account in our study.  There are
$3.1\times10^6$\ dimuon candidate events that passed the third-level
trigger requirements.

Since the average energy deposition of a muon passing through the
calorimeter system into the muon chambers was 1.4~GeV, we required
that all muon candidates have $P_T>1.4$~\GeVc; however, more stringent
criteria on the $P_T$\ of the muon candidates were imposed by the
three-level trigger system, which effectively placed a $P_T$\
threshold of approximately 2~\GeVc\ on each of the two muon
candidates.  The momentum selection of muon candidates by the trigger
enhanced the signal yield without introducing large systematic
uncertainties, since the trigger efficiency was precisely measured.

In approximately 75\%\ of our selected events, the two muon candidates
that were identified as charmonium daughters were also the muon
candidates identified by the dimuon trigger system.  In many of the
remaining events, an additional muon candidate in the event satisfied
the dimuon trigger requirements.  We included these ``volunteers'' in
our analysis in order to maximize the sensitivity of the data sample.

\subsection{Primary-Vertex and Charged-Particle Reconstruction}

We first identified the location of the \pbarp\ interaction vertex or
vertices using the observed tracks reconstructed in the VTX detector.
These tracks, when projected back to the beam axis, determined the
longitudinal locations of these interactions.  Due to the high
instantaneous luminosities, the number of reconstructed interaction
vertices in a given event follows a Poisson distribution with a mean
of $\sim$2.5.  We chose as the longitudinal position of the primary
\pbarp\ collision vertex for the dimuon candidates the interaction
vertex that was closest to either one of the muon candidates'
intercepts along the beamline.  This provided a measurement of the
primary vertex position with an accuracy of 0.3~cm along the beam
direction.

The transverse position of the primary vertex was most accurately
determined by using the average beam position through the detector and
the longitudinal primary vertex position.  The average trajectory was
stable over the period that a given \pbarp\ beam was stored in the
Tevatron.  The uncertainty in the transverse position of the primary
vertex was dominated by the transverse size of the beam, which was
25~$\mu$m in both the $x$\ and $y$\ directions.

Candidate $\mu^+$, $K^+$, and $\pi^+$\ trajectories were reconstructed
in the CTC and extrapolated into the SVX detector to identify hits
associated with the given track.  We required each CTC track candidate
to be of high quality by stipulating that a candidate track have a
minimum number of hits in the CTC.  We also required that additional
SVX information consist of at least two out of a possible four hits
(for the earlier of the two running periods, at least three hits were
required).  The CTC tracks were also required to pass through most of
the active volume of the CTC; this was imposed by demanding that the
radius of exit from the CTC volume of the charged-track trajectory be
at least 110~cm.  We also required that $K^+$\ and $\pi^+$\ candidates
have a measured transverse momentum $P_T>0.4$~\GeVc\ in order that
they be reconstructed reliably.

\subsection{Event Topology Reconstruction}
\label{sect: topo reconstruction}
We performed an event reconstruction using the muon and charged-track
candidates.  The first step of this procedure was to reconstruct a
candidate charmonium decay to the dimuon final state.  The second step
was to associate additional tracks with the charmonium candidate to
form a candidate $B$-meson decay.  During this process, we made
various selection requirements, described below, to reduce the
combinatorial backgrounds and to improve the signal-to-noise ratio for
the various decay modes.

\subsubsection{Reconstruction of \Jpsi\ and \psiprime\ Decays}
In order to reduce the rate of muon candidates from background sources
such as $K$-meson decay-in-flight, we required that each muon
candidate observed in the muon chambers correspond to a CTC track
candidate to within three standard deviations of the
multiple-scattering and measurement uncertainties in both the
transverse and longitudinal planes.

\begin{figure}
\begin{center}
\leavevmode
\hbox{%
\epsfysize=2.75in
\epsffile{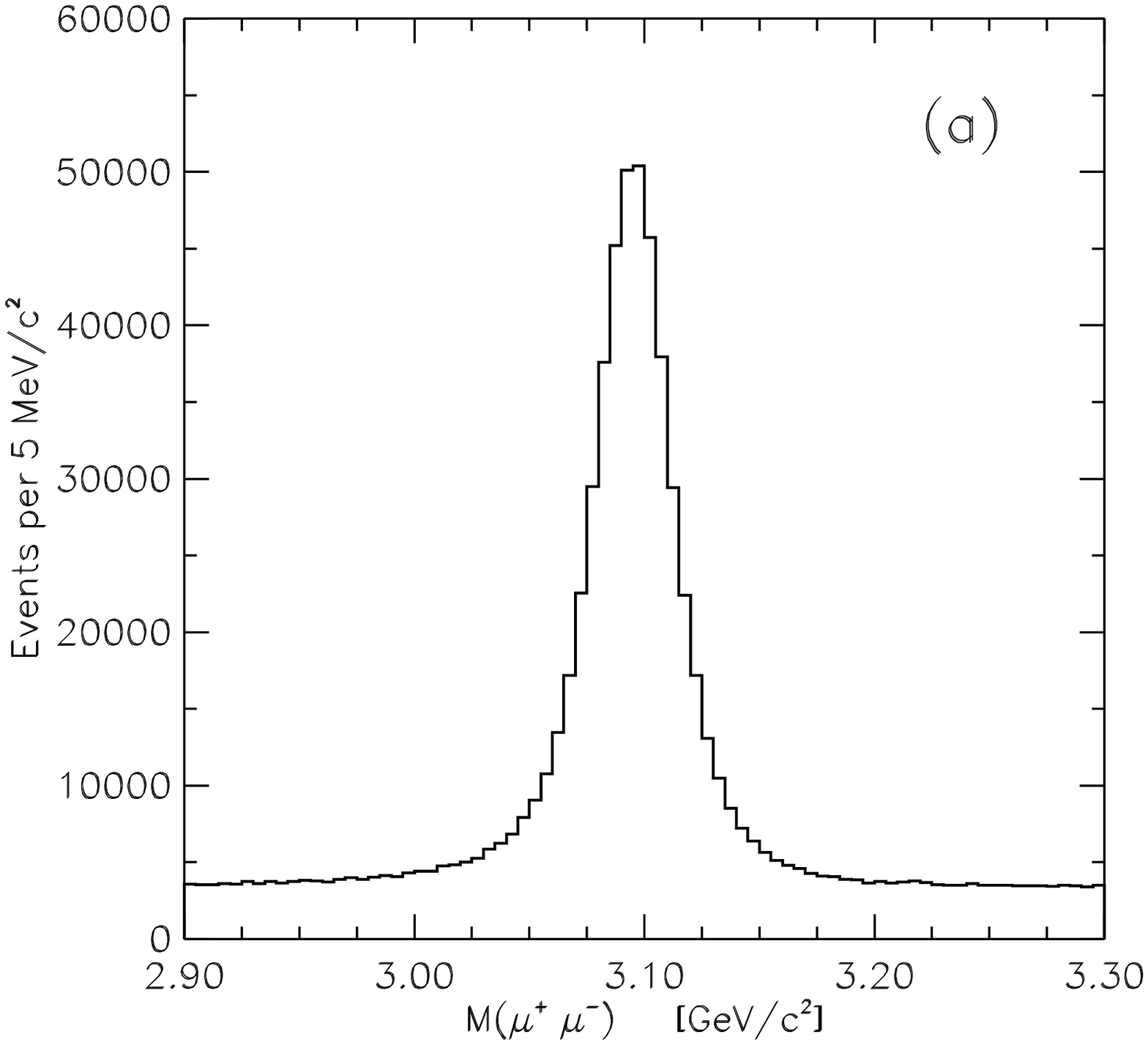}
\hspace{0.05in}
\epsfysize=2.75in
\epsffile{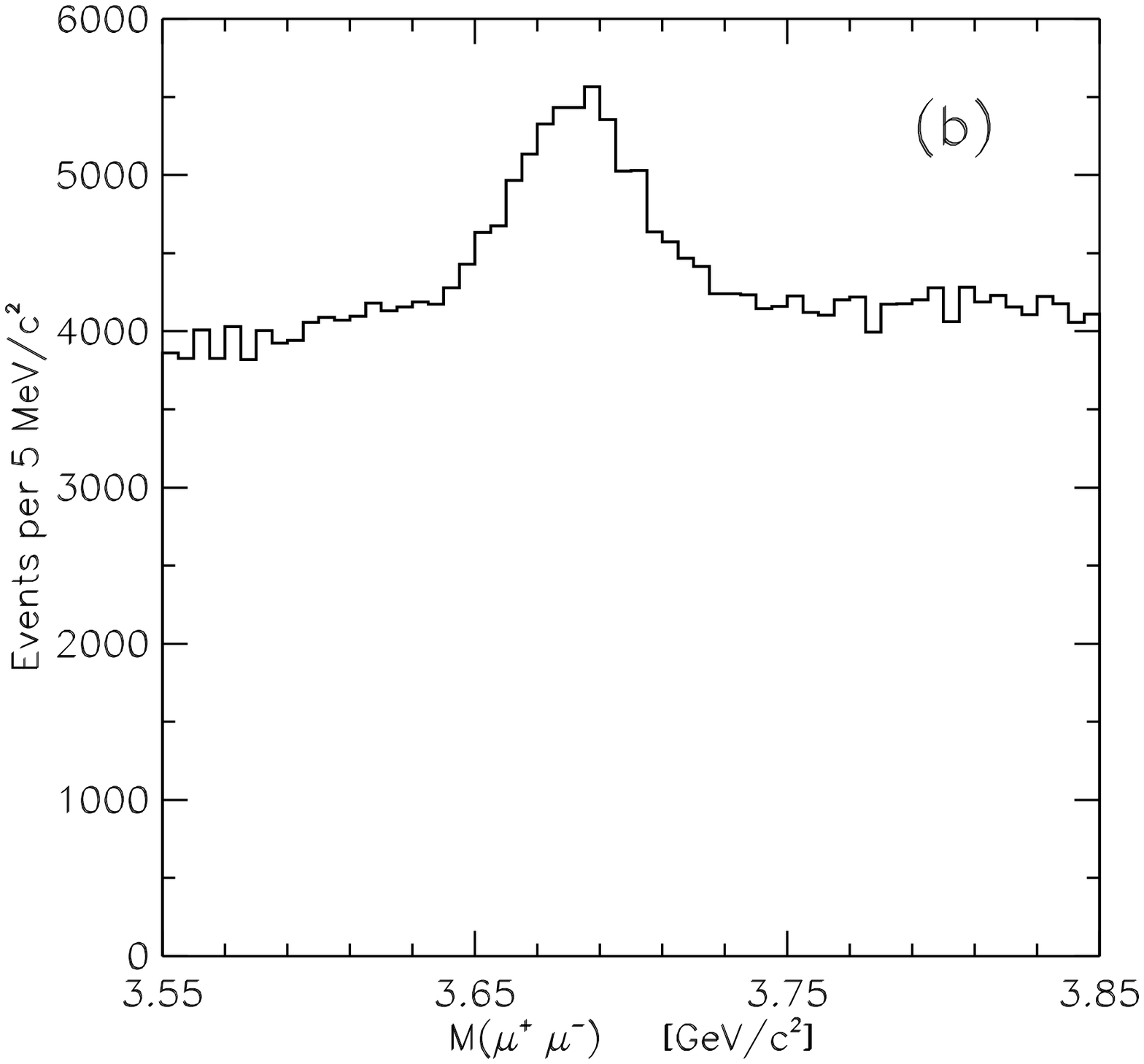}}
\end{center}
\caption{
The dimuon mass distributions for the inclusive \Jpsi\ and
\psiprime\ candidate event samples are shown in (a) and (b), respectively. 
}
\label{fig: dimuon masses}
\end{figure}

Backgrounds in the dimuon sample were further reduced by performing a
least squares fit of the CTC tracks associated with the muon
candidates under the constraint that they originate from a common
vertex.  We required that the confidence level of this fit be greater
than 0.01.  The dimuon mass distributions for the \Jpsi\ and
\psiprime\ candidates are shown in Fig.~\ref{fig: dimuon masses}.  We
fit these distributions to parametrized signal and background
lineshapes to determine that our sample consists of
$(4.39\pm0.01)\times 10^5$\ \Jpsi\ decays and
$(1.31\pm0.04)\times10^4$\ \psiprime\ decays, where we quote only the
statistical uncertainties.  Candidate dimuon events were fit with the
additional constraint that the two-muon mass equal the world-average
masses~\cite{ref: PDG} of the \Jpsi\ and \psiprime\ mesons, 3.09688
and 3.68600~\GeVcc, respectively.  We required that the confidence
level of this fit exceed $0.01$.  This requirement defined our
inclusive \Jpsi\ and \psiprime\ data sets.

\begin{figure}
\begin{center}
\leavevmode
\hbox{%
\epsfxsize=3.4in
\epsffile{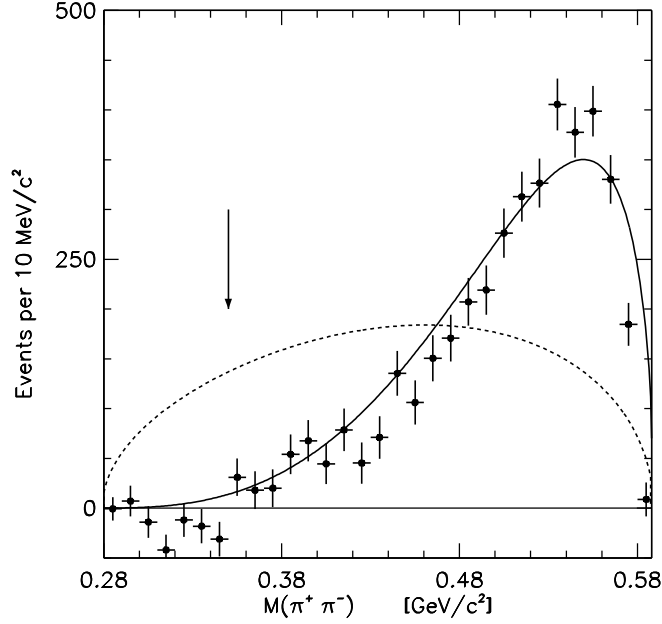}}
\end{center}
\caption{
The observed background-subtracted distribution of the dipion mass
(points) in decays of the form $\mpsiprime \rightarrow \mJpsi
\,\pi^+\,\pi^-$.  The arrow indicates the minimum mass required in the
analysis.  The solid curve represents a phenomenological prediction
due to Pham {\it et al.}~\protect\cite{ref: Pham etal} and the broken
curve describes a pure phase-space distribution.}
\label{fig: dipion mass}
\end{figure}

Candidates for the decay $\mpsiprime\rightarrow\mJpsi\,\pi^+\,\pi^-$\
were identified by combining every $\mJpsi$\ candidate identified
above with pairs of oppositely-charged track candidates that
individually had to have $P_T>0.4$~\GeVc.  The two-particle mass of
the two pion candidates was required to satisfy $0.35 < M(\pi^+\pi^-)
< 0.61$~\GeVcc.  The lower limit of this mass range was motivated by
the known dipion mass distribution for \psiprime\ decays~\cite{ref:
abrams,ref: armstrong,ref: schwinger brown cahn,ref: Pham etal}.  The
upper limit corresponded to the maximum dipion mass allowed for
\psiprime\ decays.  The efficiency of this criterion to select
\psiprime\ decays is demonstrated in Fig.~\ref{fig: dipion mass},
where we plot the two-pion mass for observed
$\mpsiprime\rightarrow\mJpsi\,\pi^+\,\pi^-$\ decays.  The background
to the \psiprime\ decays in this distribution has been removed by
performing a sideband subtraction using background events in the
$J/\psi\,\pi^+\,\pi^-$ mass distribution.  Also shown are curves
representing a phenomenological prediction~\cite{ref: Pham etal} and a
pure phase-space calculation.

\begin{figure}
\begin{center}
\leavevmode
\hbox{%
\epsfxsize=3.4in
\epsffile{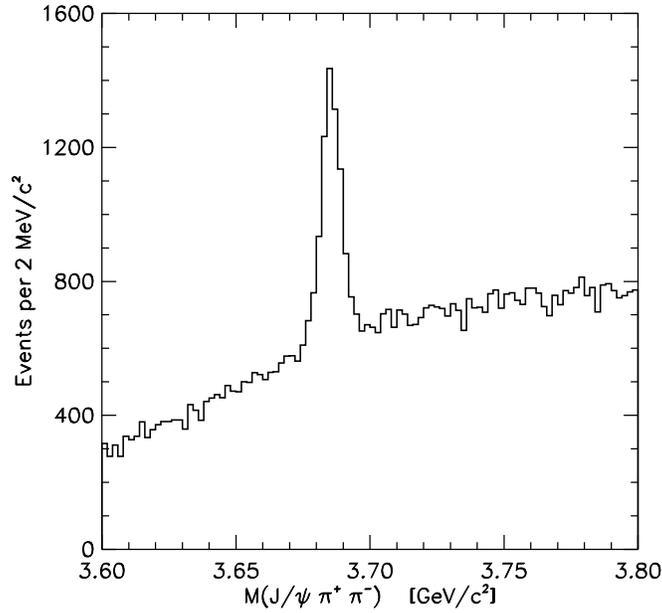}}
\end{center}
\caption{
The $\mJpsi\,\pi^+\,\pi^-$\ mass distribution for the 
$\mpsiprime$\ candidates. 
}
\label{fig: hadronic psiprime mass}
\end{figure}

To reduce combinatorial backgrounds in the $\mJpsi\,\pi^+\,\pi^-$\
candidate search further, an additional least-squares fit was
performed constraining all four tracks to emanate from a common point,
the dimuon mass to be equal to the world-average \Jpsi\ mass, and the
$\mJpsi\,\pi^+\,\pi^-$\ mass to be equal to the world-average
\psiprime\ mass.  We required that the confidence level for this fit
exceed $0.01$.  The $\mJpsi\,\pi^+\,\pi^-$\ mass distribution for all
the candidates prior to imposing the \psiprime\ mass constraint is
shown in Fig.~\ref{fig: hadronic psiprime mass} and illustrates a
narrow \psiprime\ signal of $(3.7\pm0.1)\times10^3$\ events above the
combinatorial background.

Our subsequent reconstruction identified $K^+$\ and $\pi^+$\ meson
candidates that were consistent with arising from one of the four
$B$-meson decay modes we considered in this study. 

\subsubsection{Reconstruction of $B$-Meson Exclusive Decays}
The reconstruction of the $B$-meson decay modes required the use of
selection criteria that reduced the potentially large combinatorial
background.  The most effective way of reducing these backgrounds was
to impose minimum $P_T$\ requirements on the candidate strange-meson
daughters, $P_T$\ requirements on the $B$-meson candidates, and
requirements that the decay topology be consistent with that expected
from a $B$-meson decay.  Explicit particle identification of $K^+$\
and $\pi^+$\ mesons was not employed in this analysis.

These additional requirements imposed on the candidates were also
selected to be as common as possible over the four $B$-meson decay
channels and both decay modes of the $\psi(2S)$~meson to avoid
significant systematic uncertainties arising from the estimation of
$B$-meson yields across different kinematic regions.  The kinematic
selection criteria were optimized by maximizing the quantity $N_s /
\sqrt{N_s + N_b}$, where $N_s$\ was the predicted number of signal
events based on Monte Carlo calculations and $N_b$ was the number of
observed background events in the $B$-meson mass signal region
estimated by performing extrapolations from the sideband regions.
This technique avoided the potential bias that could be introduced by
choosing selection criteria based on the number of observed candidate
signal events.

We required that each $K^+$\ candidate have $P_T>1.5$~\GeVc\ for the
reconstruction of \Bu\ candidates.  For the reconstruction of $B^0$
candidates, we required that the $K^\ast(892)^0$\ candidate have
$P_T>2.0$~\GeVc.  In the latter case, the $K^+$\ and $\pi^-$\
daughters from the \Kstarzero\ decay were required to have
$P_T>0.4$~\GeVc.  In the \Kstarzero\ reconstruction, all possible
charged-particle candidate pairs were considered with both mass
assignments.  A track-pair mass assignment was considered a candidate
\Kstarzero\ decay if the two-particle mass was within $0.0800$~\GeVcc\
of the world-average \Kstarzero\ mass of 0.8961~\GeVcc.  For a small
fraction of the track-pair candidates, it was possible that both the
$K^+\pi^-$\ and $\pi^+K^-$\ mass assignments fell within this mass
window around the \Kstarzero\ pole, resulting in double counting of
signal events.  This double counting was taken into account using a
Monte Carlo calculation described below.

To identify $B$-meson candidates, a least-squares fit was performed on
the charged-particle tracks associated with the charmonium and
strange-meson candidates, constraining them to originate from a common
decay point with the charmonium-candidate mass constraints described
above.  In addition, this fit constrained the momentum vector of the
$B$-meson candidate to be parallel to its flight path, defined by the
measured production and decay points, in the transverse plane and
required that the $B$ candidate originate from the primary interaction
vertex.  We required the confidence level of this fit to exceed
$0.01$.  We also required that the transverse momentum of the \Bu\
candidates be greater than 6.0~\GeVc\ and the transverse momentum of
the \Bd\ candidates be greater than 9.0~\GeVc.  These different
$B$-meson $P_T$\ selection criteria constitute the largest difference
in the kinematic requirements between decay modes involving \Bu\ and
\Bd\ final states and are a result of the different levels of
combinatorial background in the \psiprime\ final states.

\begin{figure}
\begin{center}
\leavevmode
\hbox{%
\epsfxsize=3.4in
\epsffile{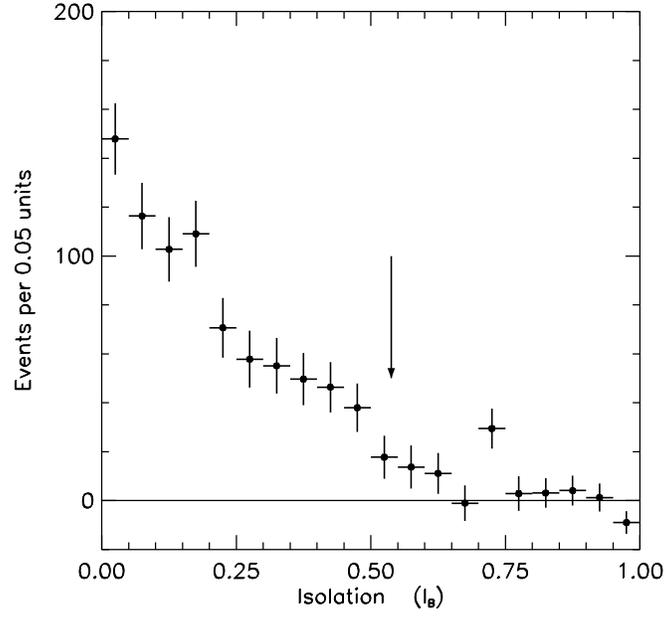}}
\end{center}
\caption{
The distribution of the isolation variable $I_B$ for candidate
$B^+\rightarrow \mJpsi \,K^+$\ decays.  A background subtraction has
been performed using the sidebands in the $J/\psi\,K^+$ mass
distribution.  The arrow represents the cut below which
candidates were accepted.}
\label{fig: isolation}
\end{figure}

Two additional criteria were imposed to reduce combinatorial
backgrounds further.  In the fragmentation of $b$ quarks into $B$
mesons, the meson typically carries most of the energy of the quark
created in the hard scattering interaction~\cite{ref: Peterson}.  We
exploit this fact to suppress backgrounds by defining an isolation
variable
\begin{eqnarray}
I_B \equiv {{\sum\limits_i \vec{P}_i \cdot \vec{P}_B }\over{|\vec{P}_B|^2}},
\end{eqnarray}
where the sum is over charged particles with momentum vectors
$\vec{P}_i$, contained within a cone in $\eta$-$\phi$ space of radius
$R\equiv\sqrt{(\Delta\phi)^2 + (\Delta\eta)^2}=1.0$\ about an axis
defined by the direction of the $B$-candidate momentum $\vec{P}_B$.
Track candidates belonging to the $B$-meson candidate were not
included in this sum.  In order to avoid including charged particles
that resulted from interactions in the \pbarp\ collision not
associated with the $B$-meson candidate, we made the sum over only
those charged-particle tracks that passed within 5~cm along the $z$
axis of the primary interaction location.  We expect $B$-meson decays
to have relatively small values of $I_B$, and therefore imposed the
requirement $I_B < 0.54$, which resulted from the optimization
procedure.  We show in Fig.~\ref{fig: isolation} the distribution of
$I_B$\ for $B^+\rightarrow \mJpsi \, K^+$\ candidate decays, after
statistically subtracting the combinatorial background under the
$B^+$\ decay signal using background events in the $\mJpsi \, K^+$\
mass-sideband regions.  This illustrates that the $B$-meson decays are
efficiently identified by this requirement.

As a final $B$-meson selection requirement, we exploited the
relatively long lifetimes of $B$~mesons and the excellent secondary
vertex resolution of the CDF detector to reject those events that have
short decay lengths~\cite{ref: Run 1 B lifetime}.  We measured the
proper decay length for each decay candidate,
\begin{eqnarray}
c\tau_B \equiv \frac{\vec{P}_T\cdot\vec{x}_T}{P^2_T}\ m_B,
\end{eqnarray}
where $m_B$\ is the mass of the $B$-meson candidate, $\vec{x}_T$\ is
the flight path measured in the transverse plane, and $\vec{P}_T$\ is
the $B$-meson vector transverse momentum.  The $c\tau_B$ resolution
depended primarily on the number of track candidates possessing SVX
hit information.  We required that $c\tau_B>100$~$\mu$m.

\subsection{$B$-Meson Signals}
\begin{figure}
\begin{center}
\leavevmode
\hbox{%
\epsfxsize=3.2in
\epsffile{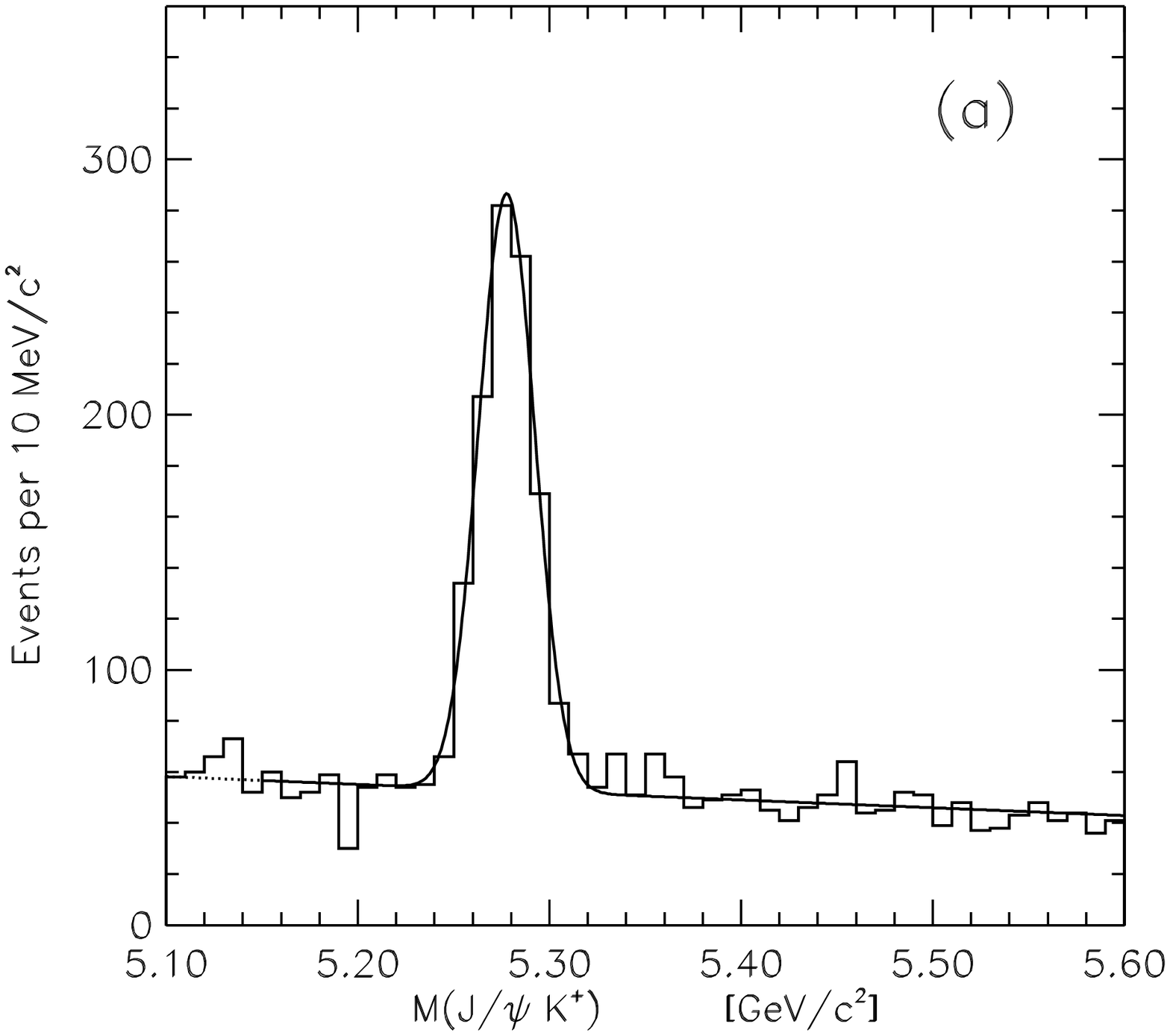}}
\vskip 0.5in
\hbox{%
\epsfxsize=3.2in
\epsffile{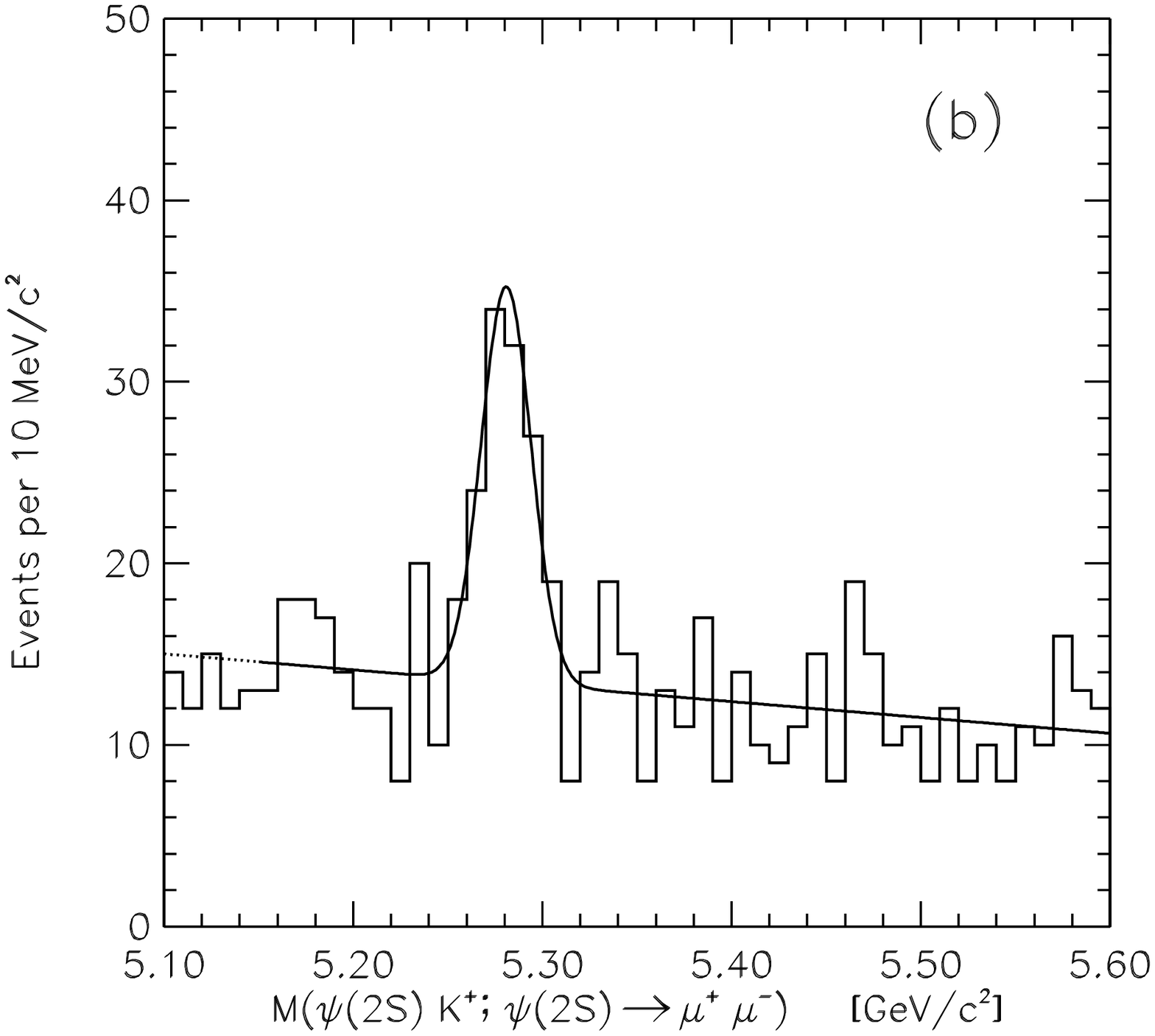}
\hspace{0.2in}
\epsfxsize=3.2in
\epsffile{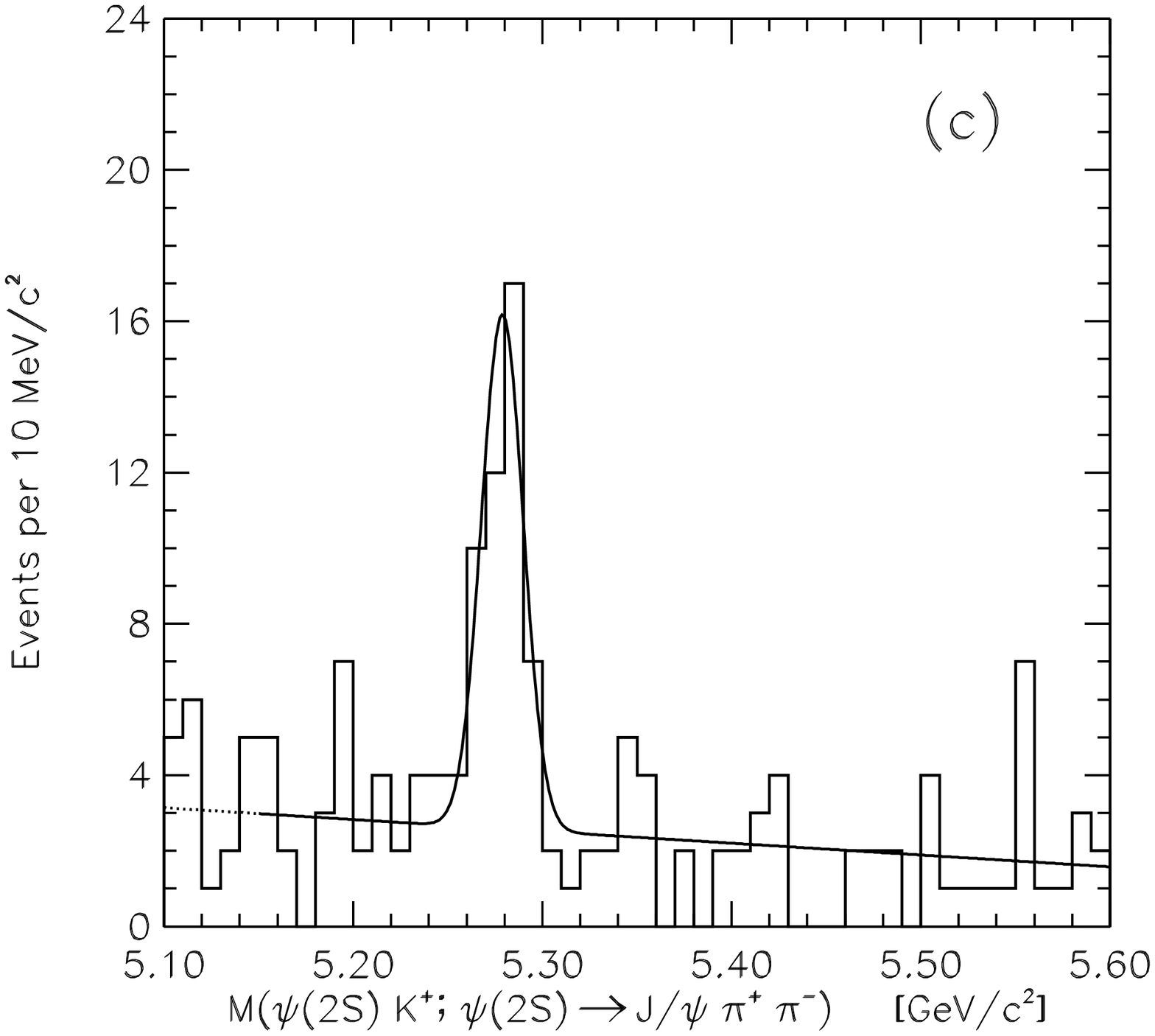}}
\end{center}
\caption{
The $\mJpsi \,K^+$\ mass distribution is shown in (a).
The $\mpsiprime \,K^+$\ mass distributions for the 
two modes $\mpsiprime\rightarrow\mu^+\,\mu^-$\ and 
$\mpsiprime\rightarrow \mJpsi\,\pi^+\,\pi^-$\ are shown in (b) and (c),
respectively.  The curves are the results of a fit described in the text.
}
\label{fig: B+ meson yields}
\end{figure}

\begin{figure}
\begin{center}
\leavevmode
\hbox{%
\epsfxsize=3.2in
\epsffile{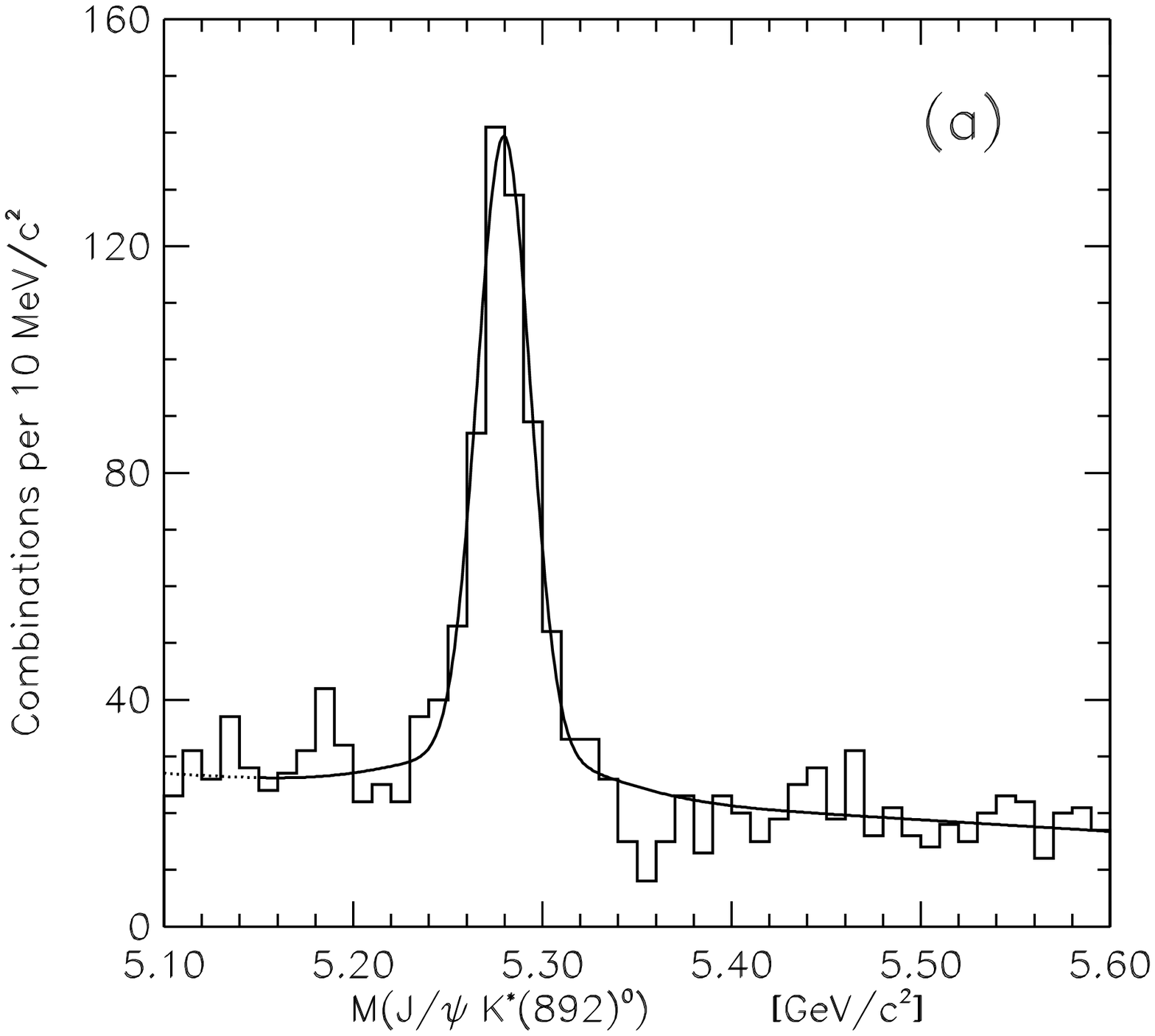}}
\vskip 0.5in
\hbox{%
\epsfxsize=3.2in
\epsffile{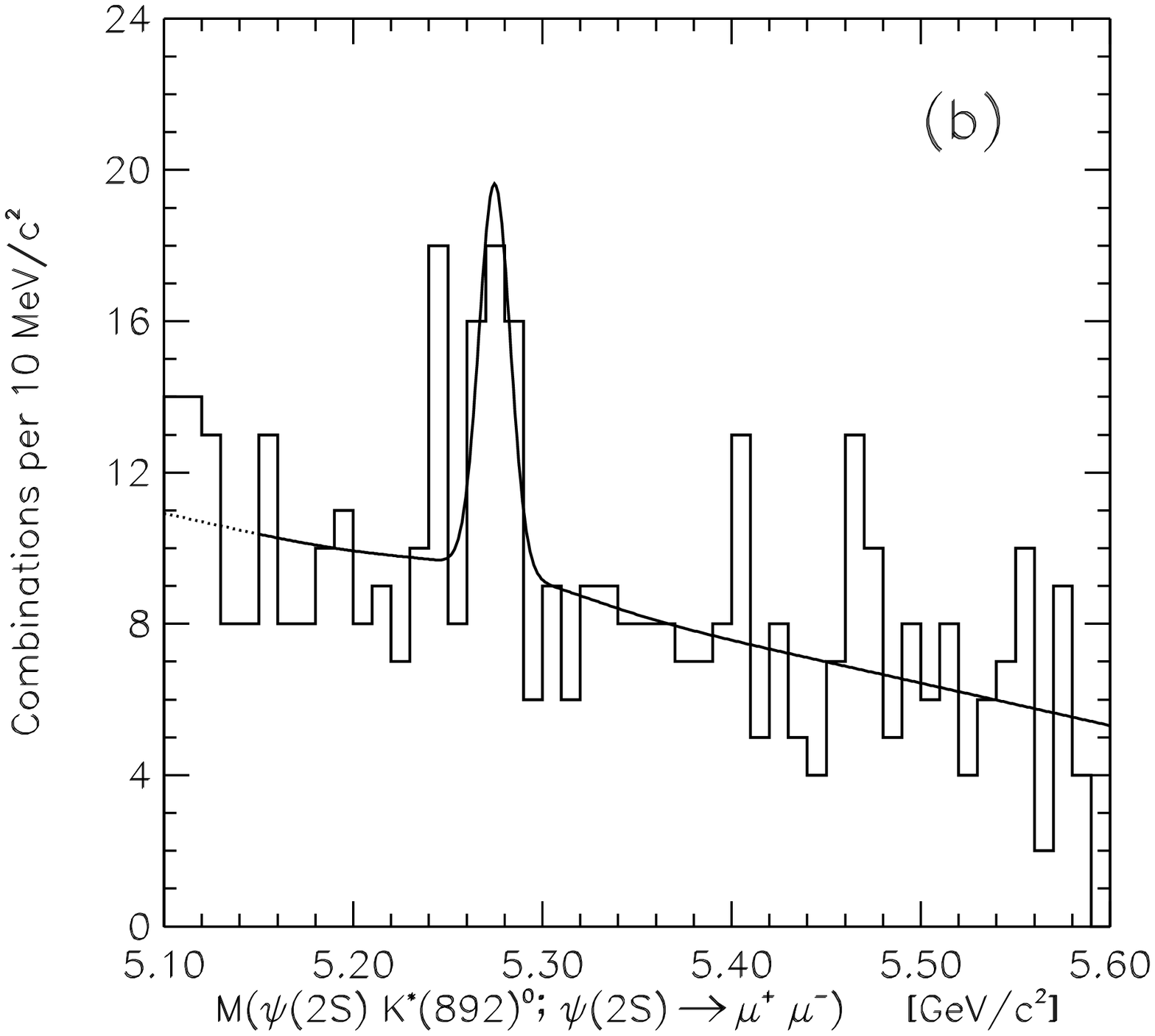}
\hspace{0.2in}
\epsfxsize=3.2in
\epsffile{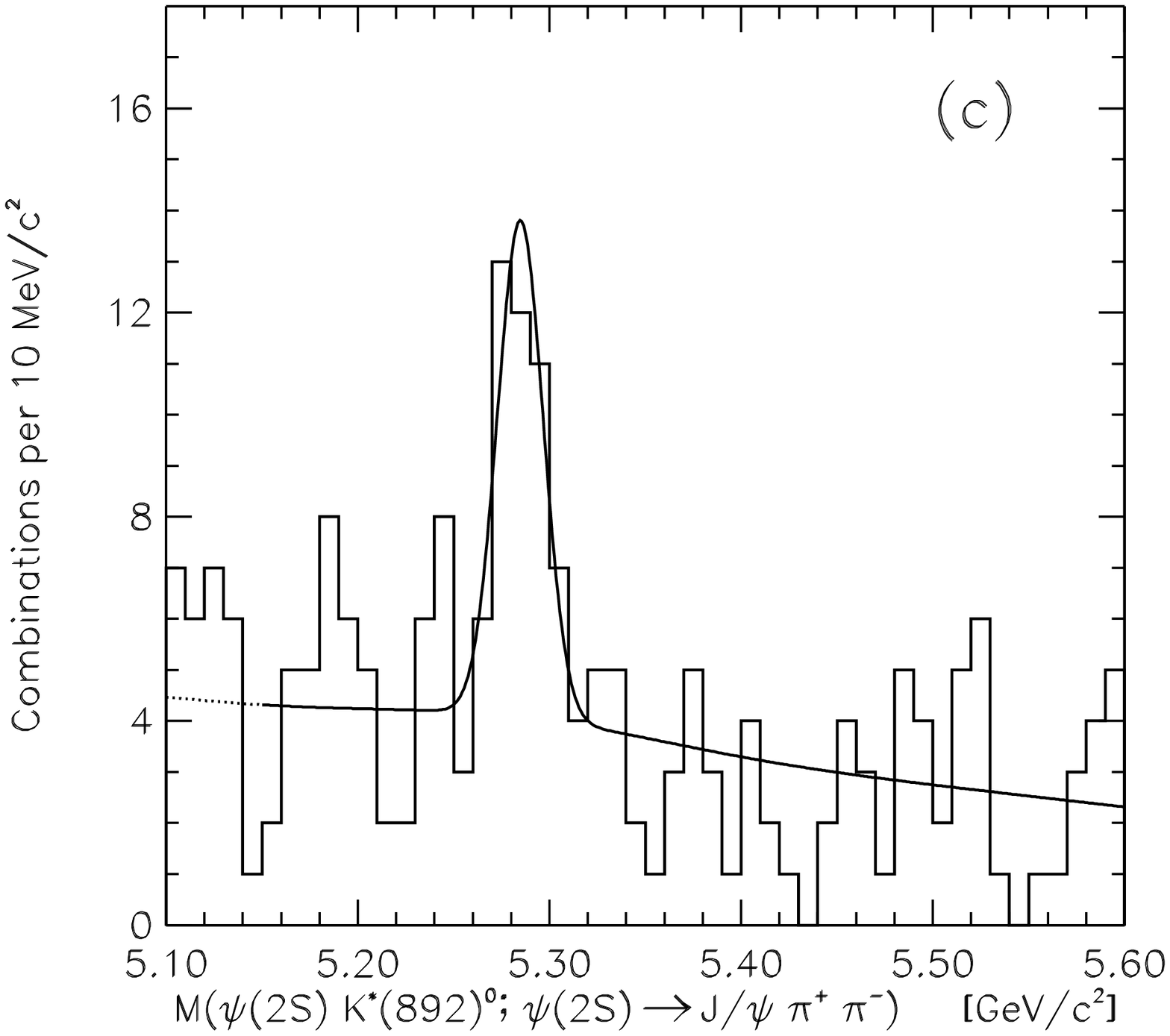}}
\end{center}
\caption{
The $\mJpsi \,\mKstarzero$\ mass distribution is shown in (a).
The $\mpsiprime \,\mKstarzero$\ mass distributions for the 
two modes $\mpsiprime\rightarrow\mu^+\,\mu^-$\ and 
$\mpsiprime\rightarrow \mJpsi\,\pi^+\,\pi^-$\ are shown in (b) and (c),
respectively.  The curves are the results of a fit described in the text.
}
\label{fig: B0 meson yields}
\end{figure}

The candidate mass distributions for the $B^+\rightarrow\mJpsi\, K^+$\
and $B^+\rightarrow \mpsiprime \, K^+$\ decays are shown in
Fig.~\ref{fig: B+ meson yields}.  The mass distributions for the
$B^0\rightarrow\mJpsi \, \mKstarzero$\ and $B^0\rightarrow \mpsiprime
\, \mKstarzero$\ decays are shown in Fig.~\ref{fig: B0 meson yields}.
We see signal peaks at the \Bu\ and \Bd\ masses in all six decay
modes.  The distributions describing the $\psi(2S)$ final states
constitute the first observations of these modes in $p\bar{p}$
collisions.  To obtain the yield of $B^+$\ candidates, each
distribution was fit to a Gaussian signal lineshape with a linear
background parametrization using a binned maximum likelihood
technique.  The fits were performed over the $B$-meson candidate mass
region from 5.15~\GeVcc\ to 5.60~\GeVcc.  The lower edge of this range
was chosen to avoid possible biases from incompletely reconstructed
$B$-meson decay modes where one or more decay daughters went
undetected.

For the \Bd\ candidates, a single Gaussian lineshape did not
accurately describe the signal due to ambiguities in the \Kstarzero\
daughter ($K^+$\ and $\pi^-$\ mesons) mass assignments in
approximately 25\% of the signal events.  To correct for this effect,
we used a Monte Carlo calculation to determine the lineshape for the
correct and incorrect mass assignments and found that each of these
was accurately described with a Gaussian parametrization centered on
the nominal $B^0$ mass.  The width of the mass distribution arising
from the wrong mass assignment was 4-6 times broader than the width of
the mass distribution associated with the right mass assignment,
although the parametrization differed for the three different $B^0$\
decay modes. We list the ratio of the peak amplitudes of the wrong
versus right mass combinations and the ratio of the widths of these
two Gaussians in Table~\ref{tab: Gaussian ratios}.  The signal yields,
masses, and resolutions of the $B^+$ and $B^0$ candidates are
summarized in Table~\ref{tab: signal fit params}.

\section{Acceptance and Efficiency Measurements}
\label{sect: efficiency corrections}
We used a Monte Carlo calculation of $b$-quark production and
$B$-meson decay, followed by a detailed detector simulation to study
and measure the kinematic and geometric acceptances for each decay
mode.  We used data to estimate the remaining efficiencies associated
with the reconstruction algorithms and the event selection criteria.

An advantage of measuring ratios of branching fractions for similar
decay modes is that many of the acceptances and efficiencies cancel in
the numerator and denominator; however, there are several effects that
do not cancel completely and have associated with them systematic
uncertainties that differ depending on which channels are being
compared.  Examples of these include the effect of the decay-length
requirement and the polarization of the $B$-meson decay to
vector-vector final states.  We have taken these into account and
arrived at separate estimates of systematic uncertainties for each of
the six ratios that were measured.

\subsection{Acceptance and Trigger Efficiency Measurements}
The Monte Carlo calculation used a model for $b$-quark production
based on a next-to-leading-order QCD calculation~\cite{ref: NDE
spectrum}.  This calculation employed the Martin-Roberts-Stirling set
D0 (MRSD0) parton distribution
functions~\cite{ref: MRS D0}\ to model the kinematics of the
initial-state partons; a value for the $b$-quark mass of $m_b =
4.75$~GeV/$c^2$; and a renormalization scale of
$\mu=\mu_0\equiv\sqrt{m_b^2 + k_T^2}$, where $k_T$\ is the momentum of
the $b$\ quark in the plane transverse to the directions of the
incoming protons.  We generated $b$~quarks in the rapidity interval
$|y_b|<1.1$\ and with $k_T>5.0$~\GeVc.  These generator limits were
chosen so that there were no biases in the $B$-meson kinematic
distributions after the application of the kinematic cuts used in this
analysis.  The $b$~quarks were fragmented into $B$~mesons according to
the Peterson fragmentation function~\cite{ref: Peterson}\ with the
parameter $\epsilon_b$\ defined to be 0.006~\cite{ref: epsilon}.

The generated $B$~mesons were decayed into the various final states
using two-body decay kinematics governed by the world-average masses
and lifetimes of the daughter particles~\cite{ref: PDG}.  For the
$\psi(2S)\to J/\psi\,\pi^+\,\pi^-$ mode, the decay matrix element used
was the Pham {\it et al.} model~\cite{ref: Pham etal,ref: coffman}, a
parametrization of which is shown in Fig.~\ref{fig: dipion mass}.

For the \Jpsi\,\Kstarzero\ and \psiprime\,\Kstarzero\ decays, which
involve two vector mesons in the final state, we used the
world-average longitudinal polarization for the \Jpsi\Kstarzero\
decay, $\Gamma_L/\Gamma = 0.78\pm0.07$~\cite{ref: Browder etal}.  The
polarization has not been measured in $B^0\to\psi(2S)\,K^*(892)^0$
decays; therefore, we use the value of $\Gamma_L/\Gamma$ for the
$B^0\to J/\psi\,K^*(892)^0$ decay but double the uncertainty to
$\pm0.14$.

The resulting generated events were passed through a detailed
simulation of the CDF detector that took into account
decays-in-flight, the geometry of all the subdetector elements, the
interaction of the charged particles with the material in the
detector, the resolution of the different tracking elements, and the
efficiency of the trigger. 

We used a parametrization of the muon-$P_T$ dependence of our first
and second-level trigger system to determine the trigger efficiency
for the various $B$-meson decays in which the muons that resulted from
the charmonium decay were also identified by the trigger's dimuon
selection criteria.  In the remaining events, which corresponded to
approximately 25\%\ of the selected event sample, an additional muon
candidate in the event often satisfied the dimuon trigger
requirements.  This led to a $\sim$5\%\ uncertainty in the topology
dependence of the relative trigger efficiency.  We included this as a
separate systematic uncertainty in our branching-fraction
measurements.

The resulting geometric acceptances for the different decay modes
are presented in Table~\ref{tab: acceptances}, where we list the
fraction of decays expected to contribute to the observed signal peak.

\subsection{Reconstruction Efficiencies}
\label{sect: recon efficiencies}
The efficiencies of the subdetectors and reconstruction algorithms
used in this analysis to identify charged-particle candidates and
reject backgrounds were separately estimated using information from
the collected data together with Monte Carlo calculations.  The use of
data that properly represented the running conditions experienced at
the Tevatron Collider was particularly important in cases where these
efficiencies were expected to depend on the instantaneous luminosity.

\subsubsection{Track Reconstruction Efficiency}
Because of the different number of charged particles used in the
reconstruction of each of the exclusive decay modes in this analysis,
we had to estimate the efficiency of the track reconstruction
algorithms when reconstructing three, four, five, or six-prong decays.
The large number of charged particles associated with an interaction
producing a $B$~meson (typically $\sim$40 charged particles in the
fiducial region of the CTC) and the large number of simultaneous
interactions created very high hit occupancies in the innermost layers
of the CTC, reducing their effectiveness.

We measured the efficiency of the CTC track-finding algorithm by
embedding the wire hits from one or two Monte Carlo charged-particle
tracks into a set of data events identified as having a \Jpsi\
candidate.  This event sample was selected to be representative of the
inclusive \Jpsi\ and \psiprime\ data set, taking into account
variations in detector configuration and instantaneous luminosity.
Our embedding procedure used the hit detection efficiencies and
resolutions observed in the data to simulate the response of the CTC.
We then used the standard track reconstruction algorithms to seek a
reconstructed track helix that formed a match with the embedded
particle's trajectory.  The track matching criterion was imposed on a
$\chi^2$ variable with five degrees of freedom that accounted for
parameter correlations and measurement uncertainties by employing in
its definition the covariance matrix of the reconstructed track helix.
For an embedded Monte Carlo track in a data event, $\chi^2$
values were computed for all of the reconstructed tracks in the event.
A track was deemed to be a match if it had the lowest $\chi^2$ of all
tracks in the event and if this $\chi^2$ value was $<$500, a highly
efficient requirement.

We measured the efficiency for reconstructing a $\pi^\pm$\ meson as a
function of the meson's $P_T$, shown in Fig.~\ref{fig: track
efficiency}, and as a function of several other variables~\cite{ref:
thesis}.  We found the average efficiency for $\pi^\pm$\ mesons to be
$0.928\pm0.003\pm0.026$\ for meson $P_T>0.4$~\GeVc, where the quoted
uncertainties are statistical and systematic, respectively.  Because
of the use of the decay mode
$\mpsiprime\rightarrow\mJpsi\,\pi^+\,\pi^-$, which introduced two
additional charged tracks, we also repeated this procedure by
embedding pairs of charged tracks in each real event.  We found the
efficiency for reconstructing the two daughter pion tracks in the
decay $\mpsiprime\rightarrow\mJpsi\,\pi^+\,\pi^-$\ to be
$0.881\pm0.005\pm0.043$.  The systematic uncertainties in these two
measurements are largely correlated because they were dominated by the
uncertainty in the CTC hit efficiencies used in the track embedding
procedure.  Since the square of the single-track efficiency is
substantially smaller than the two-track reconstruction efficiency, we
concluded that the track-finding efficiencies of several charged
particles in a single event were correlated, an effect that was taken
into account in our subsequent efficiency calculations.

\subsubsection{Requirements on the Fits to Decay Topologies}
We measured the efficiencies of the constrained-fit confidence-level
criteria for the different decay topologies using the observed \Jpsi\
and \psiprime\ signal yields before and after making the
confidence-level requirements discussed in Sec.~\ref{sect: topo
reconstruction}.  These comparisons resulted in efficiencies for the
constrained fits to the $\mJpsi\rightarrow\mu^+\,\mu^-$\ and the
$\mpsiprime\rightarrow\mu^+\,\mu^-$\ decay hypotheses that were equal
within uncertainties.  The efficiency for satisfying the
confidence-level requirement on the fit employing a vertex constraint
was $0.967\pm0.003$\ and the efficiency for the fit employing the
additional charmonium mass constraint was $0.963\pm0.002$.  Since
these efficiencies do not depend on the charmonium parent, they cancel
in the measurement of the ratios of branching fractions.  The
uncertainty in these measurements arises from the finite statistics of
the two charmonium samples.

The third charmonium topology,
$\mpsiprime\rightarrow\mJpsi\,\pi^+\,\pi^-$, includes two low-momentum
charged pions in the decay vertex in addition to the muons.  We
measured the extra inefficiency caused by the vertex-constrained fit
to this decay by measuring the loss of \psiprime\ signal events when
making the confidence-level requirement.  This resulted in a
correction factor of $0.834\pm0.039$ for the efficiency of the
requirement on the vertex fit alone and a factor of $0.945\pm0.031$
corresponding to the additional mass-constrained fit requirement.

\subsubsection{Decay-Length Requirement}
The efficiencies of the $B$-meson decay-length requirement on the
different decay modes discussed in Sec.~\ref{sect: topo
reconstruction} were measured by modeling the decay-length resolution,
which was determined by the number of tracks reconstructed in the SVX
detector, and using the world-average lifetimes for the \Bu\ and \Bd\
mesons~\cite{ref: PDG}.  We determined the decay-length resolutions
for events with different numbers of SVX tracks using $B$-meson
candidate events outside of the signal mass region.  We then used the
Monte Carlo calculation and detector simulation to estimate the
expected frequencies of events with different numbers of SVX tracks.
These were then convolved together with the expected exponential
decay-length distributions for each $B$-meson species to obtain a
prediction for the observed decay-length distributions.

The efficiencies of the decay-length requirement were $\sim$0.75 and
varied only slightly between decay modes.  The uncertainties in these
efficiencies were dominated by uncertainties in the world-average
$B$-meson lifetimes~\cite{ref: PDG} and in the number of candidate
daughter tracks with SVX hit information.

\subsubsection{Isolation Requirement}
The $B$-meson isolation criterion $I_B$ discussed in Sec.~\ref{sect:
topo reconstruction} rejected combinatorial background.  The
efficiency of this selection requirement is $0.928\pm0.054$, which we
measured by estimating the loss of $B$-meson decays when applying this
criterion for each decay mode.  We note, for example, that we improve
the signal-to-noise ratio in the $\mJpsi K^+$\ channel by a factor of
three when applying this criterion.

We assume that the isolation criterion is independent of the
final-state topology.  We do not expect this isolation requirement to
depend on the specific $B$-meson final state, since it was defined
using only the $B$-meson momentum and the momentum of the charged
particles produced in the production and fragmentation of the
$b$~quark.  We verified this by measuring the efficiencies
independently in each channel, but the statistical power of this check
is limited.

\subsection{Relative Acceptance and Efficiency Corrections}
For the purpose of determining relative branching fractions, the
geometric and kinematic acceptance results listed in Table~\ref{tab:
acceptances} were combined into twelve acceptance ratios involving the
six reconstructed channels.  Table~\ref{tab:rel_geom_eff_syst}
provides a summary of these relative acceptances and their systematic
uncertainties.  The uncertainties include contributions from the
following sources: finite Monte Carlo statistics, variations in the
Monte Carlo $P_T$ spectrum with changes in the renormalization scale
and $b$-quark mass, modeling effects in the simulation of the
second-level trigger, variations in the assumed longitudinal
polarization fractions for the decays with vector-vector final states,
and the uncertainty in the CDF detector simulation.

Table~\ref{tab:rel_all_eff_syst} is similarly structured to present
the ratios of the products of the remaining efficiencies described in
Sec.~\ref{sect: recon efficiencies}, namely those associated with
track reconstruction, the constrained-fit confidence-level criteria,
and the proper decay-length requirement.

\section{Results}
\label{sect: results}
We present our results as six ratios of acceptance-corrected rates of
$B$-meson decays into the four channels.  In ratios involving
$\psi(2S)$ mesons, the results for the two $\psi(2S)$ decay modes were
combined, taking into account the \psiprime\ branching fractions.  The
observed numbers of signal events for each decay, listed in
Table~\ref{tab: signal fit params}, were divided by the acceptance and
reconstruction efficiencies listed in
Tables~\ref{tab:rel_geom_eff_syst}\ and \ref{tab:rel_all_eff_syst},
respectively.  The event rates were also corrected for the
daughter-meson branching fractions listed in Table~\ref{tab: secondary
BR}.  Additional systematic uncertainties of 4\% and 2\% were applied
to the branching-fraction ratio calculations to account for
uncertainties in the fitting technique used to estimate event yields
and the lack of cancellation of the efficiencies of the $B$-candidate
constrained-fit criteria, respectively.  When we form the ratios of
acceptance-corrected event rates, the $b$-quark production cross
section, time-integrated luminosity, and common efficiencies divide
out of the calculations.  For both the $B^+$ and $B^0$-meson cases, we
verified that the event rates for the two different \psiprime\ decay
modes were consistent after taking into account the differences in
acceptance and reconstruction efficiencies.

The measured quantities are the ratios of the product of $b$-quark
fragmentation fractions and the $B$-meson branching fractions into the
specific final state.  Thus, our measurements can be written as shown
in Table~\ref{tab: final results}, where the first uncertainty is
statistical and the second uncertainty is the sum in quadrature of
uncertainties in acceptance, efficiency, and the charmonium daughter
branching fractions.  This convention will be employed below unless
otherwise noted.  The ratio in Table~\ref{tab: final results}\
involving exclusively the \Jpsi\ decay mode has been measured
previously with a subset of these data~\cite{ref: Run 1A BR}.  The
present measurement supersedes it.

The results show that the rates of $B$-meson decays to the two
\psiprime\ final states are approximately 50\%\ of the rates of the
analogous decays to the \Jpsi\ final states.  In addition, we note
that the relative rates of vector-vector decays to vector-pseudoscalar
decays for the \psiprime\ and \Jpsi\ modes are the same to within
relatively large statistical uncertainties.  If we make the
assumption~\cite{ref: Run 1A BR,ref: PDG} of equal fragmentation
fractions, $f_u=f_d$, the vector-vector to vector-pseudoscalar decay
rate for the \Jpsi\ final states,
\begin{eqnarray}
{{
\mBR(B^0 \rightarrow J/\psi \, \mKstarzero)
               }\over{
\mBR(B^+\rightarrow J/\psi \, K^{+})
						}}
&=&    1.76\pm0.14\pm0.15,
\end{eqnarray}
is now the most precise single measurement of this quantity and
provides a constraint on theoretical calculations that attempt to
describe both this branching-fraction ratio and the longitudinal
polarization of the vector-vector decay~\cite{ref: J/psi
polarization,ref: gourdin_etal_aleksan_etal,ref: kamal_santra}.  The
analogous result for $\psi(2S)$ final states may be similarly obtained
from Table~\ref{tab: final results}.

\subsection{Comparison with Phenomenological Models}
We have compared the results of our measurements to two
phenomenological predictions for these ratios based on the
factorization hypothesis~\cite{ref: BSW semileptonic,ref: BSW
hadronic}.  In this class of models, the ratios of branching fractions
of \Jpsi\ and \psiprime\ decay modes depend largely on the strong
interaction effects in the final state of each decay, which are
modeled by sets of form factors specific to each decay.  We have
chosen two specific models~\cite{ref: Neubert etal,ref: Deandrea
etal}, as predictions for the branching fractions for \psiprime\ and
\Jpsi\ decays have been made using them.

The calculation by Neubert {\it et al.}~\cite{ref: Neubert etal}\ employs a
set of form factors determined using a relativistic harmonic
oscillator potential model for the meson wave functions and a dipole
$q^2$~dependence for most of the form factors, where $q^2$\ is the
square of the four-momentum exchanged between the $B$~meson and the
daughter $K$ meson.  (The generic multipole form-factor formula is
$F(q^2) = F(0) / (1 - q^2/m^2)^n$, where $m$ is the pole mass and
$n=1$ or 2 for a monopole or dipole dependence, respectively.)  The
calculation by Deandrea {\it et al.}~\cite{ref: Deandrea etal}\ determines
the form factors empirically by extracting them from semileptonic
$D$-meson decays and assuming that they have a monopole $q^2$
dependence.

\begin{figure}
\begin{center}
\leavevmode
\hbox{%
\epsfxsize=5.0in
\epsffile{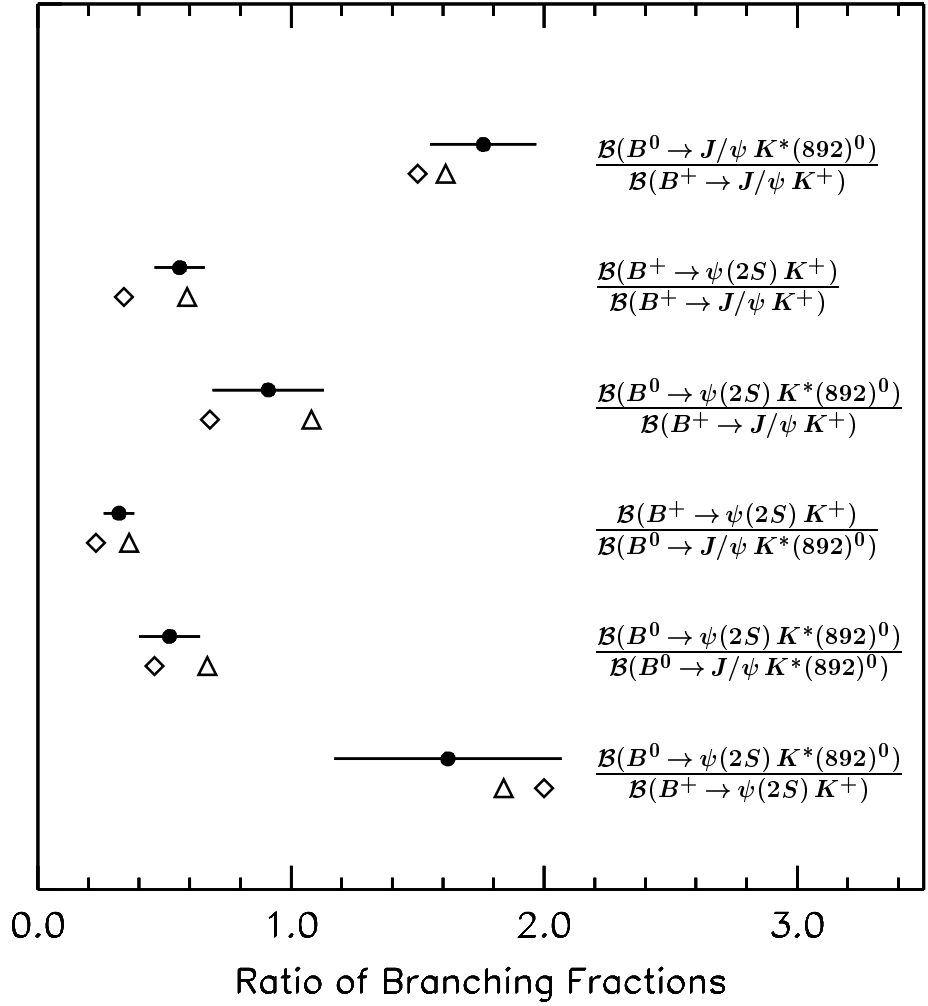}}
\end{center}
\caption
{Comparison of the measured branching-fraction ratios (filled circles)
with theoretical predictions by Neubert {\it et
al.}~\protect\cite{ref: Neubert etal} (triangles) and Deandrea {\it et
al.}~\protect\cite{ref: Deandrea etal} (diamonds).  The error bars on
the measured values represent the statistical and systematic
uncertainties added in quadrature.  In ratios involving $B^+$ and
$B^0$ mesons, $f_u = f_d$ has been assumed.}
\label{fig: theory comparison}
\end{figure}

The measured ratios of branching fractions are compared with these two
theoretical calculations in Fig.~\ref{fig: theory comparison}, where
we have assumed $f_u=f_d$.
Both models are in agreement with the data. However,
the results of Neubert \etal\ are favored overall by the data,
whereas the calculation of Deandrea \etal\ predicts a lower branching
fraction for the $B^+ \rightarrow \mpsiprime \, K^+$\ decay.

\subsection{Absolute Branching Fractions}
We can use the world-average branching fraction~\cite{ref: PDG}
\begin{eqnarray}
\label{eq: pdg bjpkst}
\mBR(B^+\rightarrow \mJpsi \, K^+) = \left( 1.01\pm0.14\right)\times 10^{-3}
\end{eqnarray}
and the assumption $f_u = f_d$ to convert our measurements of
branching-fraction ratios into absolute branching-fraction
measurements for the three other decay modes.  There are no
correlations between our data and the world-average value in
Eq.~(\ref{eq: pdg bjpkst}), making the determination of the resulting
uncertainties straightforward.  We note that the world-average value
is based on the assumption of equal $f_u$ and $f_d$ fragmentation
fractions of $b$~quarks produced in the decay of the $\Upsilon(4S)$\
meson.

The derived branching fractions are therefore
\begin{eqnarray}
\mBR(B^+ \rightarrow \mpsiprime \,K^+) &=& (0.56\pm0.08\pm0.10)
	\times 10^{-3} \\
\mBR(B^0 \rightarrow \mpsiprime \,\mKstarzero) &=& (0.92\pm0.20\pm0.16)
	\times 10^{-3}\\
\mBR(B^0 \rightarrow \mJpsi \,\mKstarzero) &=& (1.78\pm0.14\pm0.29)
	\times 10^{-3}.
\end{eqnarray}
The first and second uncertainties are statistical and systematic,
respectively.  

\begin{figure}
\begin{center}
\leavevmode
\hbox{%
\epsfxsize=5.0in
\epsffile{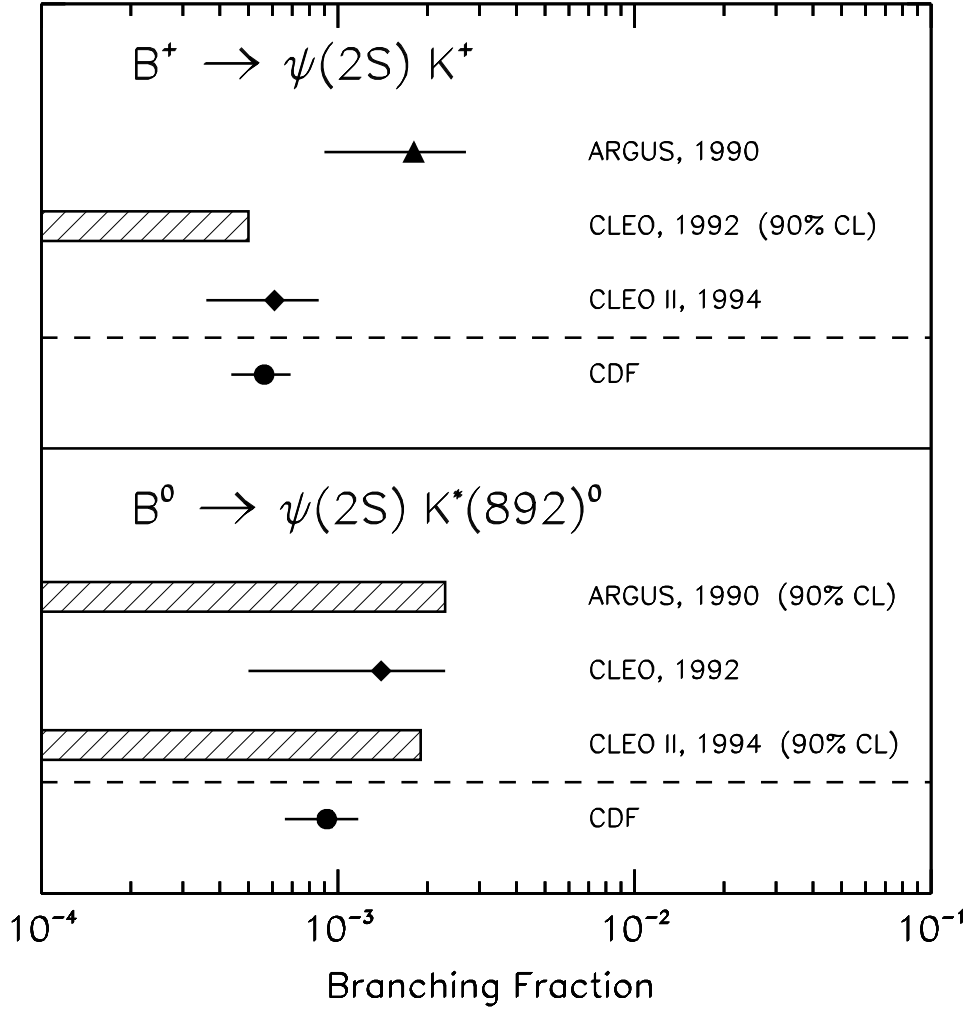}}
\end{center}
\caption
{A comparison of the derived CDF $\mBR(B^+\to\psi(2S)\,K^+)$ and
${\cal B}(B^0\to\psi(2S)\,K^*(892)^0)$ absolute branching fractions
with measurements and limits from the ARGUS~\protect\cite{ref: ARGUS
measurement}, CLEO~\protect\cite{ref: CLEO measurement}, and
CLEO~II~\protect\cite{ref: CLEO II measurement} experiments.  The
hatched bars denote 90\% confidence-level (C.L.) upper limits and the
error bars represent the statistical, systematic, and
branching-fraction uncertainties added in quadrature.}
\label{fig: previous measurements}
\end{figure}

The branching fractions for the \psiprime\ decay modes are compared
with previous measurements by the ARGUS~\cite{ref: ARGUS measurement},
CLEO~\cite{ref: CLEO measurement}, and CLEO II~\cite{ref: CLEO II
measurement}\ Collaborations in Fig.~\ref{fig: previous measurements}.
The CDF results are in agreement with these previous measurements and
have uncertainties that are approximately three times smaller.  They
represent a significant improvement in the knowledge of $B$-meson
decays to \psiprime\ final states.  The $J/\psi\,K^*(892)^0$ branching
fraction is also in good agreement with the world-average
value~\cite{ref: PDG} and a recent measurement published by the CLEO
II Collaboration~\cite{ref: Latest CLEO II measurement}.

\section{Conclusion}
\label{sect: conclusion}
We have made measurements of the rates of the exclusive decays
\begin{eqnarray}
B^+ &\rightarrow& \mpsiprime \,K^+ \nonumber \\
B^0 &\rightarrow& \mpsiprime \,K^\ast(892)^0 \\
B^+ &\rightarrow& \mJpsi \,K^+ \nonumber \\
B^0 &\rightarrow& \mJpsi \,K^\ast(892)^0. \nonumber
\end{eqnarray}

Using the observed event rates, correcting for the relative
efficiencies for these decay modes, and assuming equal fragmentation
probabilities for \Bu\ and \Bd\ mesons, we measure the following ratios of
branching fractions:
\begin{eqnarray}
{{
\mBR(B^0 \rightarrow J/\psi \,\mKstarzero)
               }\over{
\mBR(B^+\rightarrow J/\psi \,K^{+})
						}}
&=&    1.76\pm0.14\pm0.15 
\nonumber \\[0.1in]
{{
 \mBR(B^+ \rightarrow \mpsiprime \,K^{+})
               }\over{
 \mBR(B^+\rightarrow \mJpsi \,K^{+})
						}}
&=&    0.558\pm0.082\pm0.056 
\nonumber \\[0.1in]
{{
\mBR(B^0 \rightarrow \mpsiprime \,\mKstarzero)
               }\over{
\mBR(B^+\rightarrow J/\psi \,K^{+})
						}}
&=&    0.908\pm0.194\pm0.100\\[0.1in]
{{
\mBR(B^+ \rightarrow \mpsiprime\,K^+)
               }\over{
\mBR(B^0 \rightarrow \mJpsi\,\mKstarzero)
						}}
&=&    0.317\pm0.049\pm0.036 \nonumber \\[0.1in]
{{
\mBR(B^0 \rightarrow \mpsiprime\, \mKstarzero)
               }\over{
\mBR(B^0 \rightarrow \mJpsi\,\mKstarzero)
						}}
&=&    0.515\pm0.113\pm0.052 \nonumber \\[0.1in]
{{
\mBR(B^0 \rightarrow \mpsiprime\, \mKstarzero)
               }\over{
\mBR(B^+ \rightarrow \mpsiprime \, K^+)
						}}
&=&    1.62\pm0.41\pm0.19. \nonumber 
\end{eqnarray}
These ratios have been compared to two phenomenological calculations
that use the factorization hypothesis.  The calculations reproduce the
overall features of the trends observed in the data.

We have also used the world-average~\cite{ref: PDG} branching fraction
$\mBR(B^+\rightarrow \mJpsi \,K^+)$\ to determine the absolute
branching fractions
\begin{eqnarray}
\mBR(B^+ \rightarrow \mpsiprime \, K^+) &=& (0.56\pm0.08
	\pm0.10) \times 10^{-3} \nonumber \\
\mBR(B^0 \rightarrow \mpsiprime \, \mKstarzero) &=& (0.92\pm0.20
	\pm0.16) \times 10^{-3} \\
\mBR(B^0 \rightarrow \mJpsi \, \mKstarzero) &=& (1.78\pm0.14\pm
	0.29) \times 10^{-3}.
\nonumber
\end{eqnarray}
These measurements represent a significant improvement in the
knowledge of $B$-meson decay rates into $\psi(2S)$ final states and
contribute to an effective test of contemporary $B$-meson decay
models.

\section{Acknowledgments}
We thank the Fermilab staff and the technical staff at the
participating institutions for their essential contributions to this
research.  This work was supported by the U.~S.~Department of Energy
and National Science Foundation; the Natural Sciences and Engineering
Research Council of Canada; the Istituto Nazionale di Fisica Nucleare
of Italy; the Ministry of Education, Science, and Culture of Japan;
the National Science Council of the Republic of China; the A.~P.~Sloan
Foundation; the Swiss National Science Foundation; and the German
Bundesministerium f\"{u}r Bildung, Wissenschaft, Forschung, und
Technologie.

%

%
%

\newpage

\newpage

\newpage

\newpage

\begin{figure}
\begin{center}
\leavevmode
\hbox{%
\epsfxsize=3.2in
\epsffile{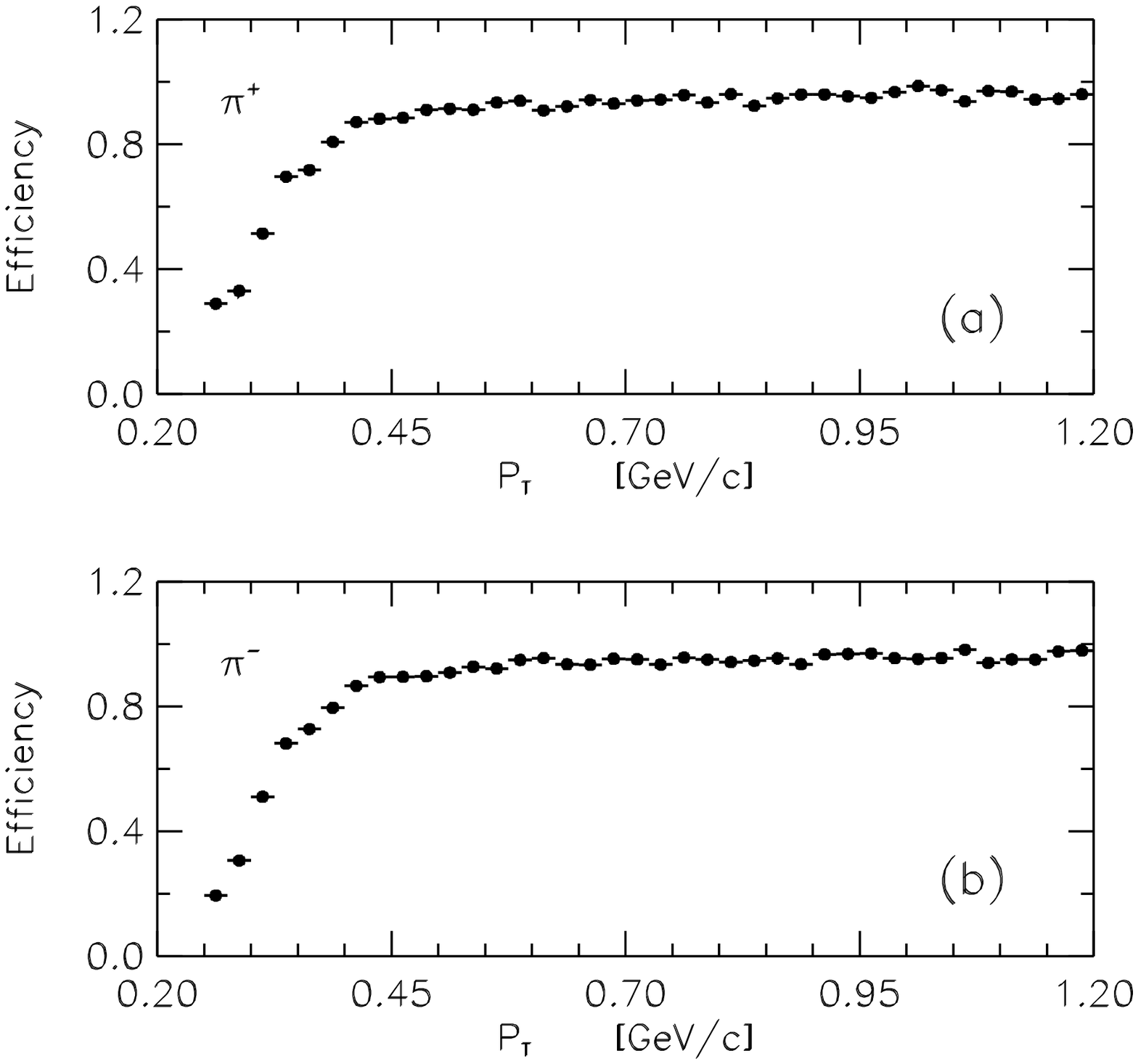}
\hspace{0.2in}
\epsfxsize=3.2in
\epsffile{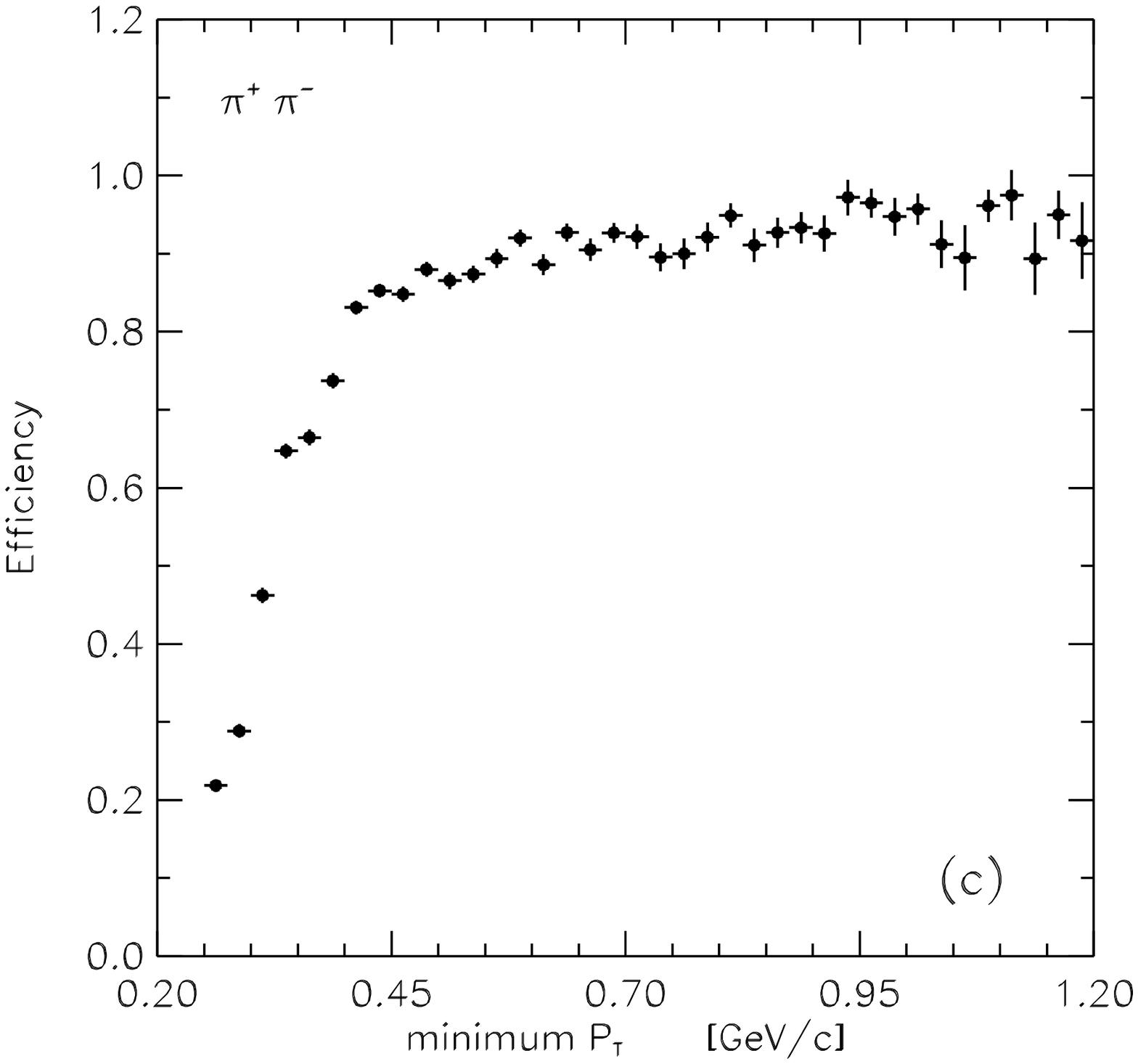}}
\end{center}
\caption{
The track reconstruction efficiency for $\pi^+$\ mesons and $\pi^-$\
mesons as a function of the meson $P_T$\ is shown in (a) and (b),
respectively.  The dipion track reconstruction efficiency as a
function of the lower momentum of the two pions is shown in (c).  }
\label{fig: track efficiency}
\end{figure}

\newpage

\newpage

\newpage


%
%
\begin{table}
\caption{
The ratios of the peak amplitudes and widths of the Gaussian
parametrizations describing the wrong to right $K$-$\pi$\
mass assignments in the \Kstarzero\ reconstruction.}
\begin{center}
\begin{tabular}{llcc}
\multicolumn{1}{c}{$B$-Meson Decay}   &   \multicolumn{1}{c}{$c\bar{c}$ Mode}
	& Amplitude  &  Width \\
&       &   Ratio        & Ratio \\
\hline
$B^0 \rightarrow \mJpsi \,\mKstarzero$ & $\mJpsi\rightarrow \mu^+\,\mu^-$ &
	0.068  & 3.6  \\
$B^0 \rightarrow \mpsiprime \,\mKstarzero$ &  $\mpsiprime\rightarrow
\mu^+\,\mu^-$ &
	0.046  & 5.8 \\
$B^0 \rightarrow \mpsiprime \,\mKstarzero$ & 
$\mpsiprime\rightarrow\mJpsi\,\pi^+\,\pi^-$ &
	0.043 & 6.3 \\
\end{tabular}
\end{center}
\label{tab: Gaussian ratios}
\end{table}

%
%
\begin{table}
\caption{The numbers of observed signal events, the fitted masses, and
the signal widths for the six different $B$-decay reconstructions.  The
listed uncertainties are statistical only.}
\begin{center}
\begin{tabular}{llccc}
\multicolumn{1}{c}{$B$-Meson Decay}   &   \multicolumn{1}{c}{$c\bar{c}$ Mode}
	&   Event Yield & Mass~[MeV/$c^2$] & Width~[MeV/$c^2$]\\
\hline
$B^+ \rightarrow \mJpsi \,K^+$     &  $\mJpsi\rightarrow \mu^+\,\mu^-$ &
          $856.7\pm38.3$ & $5278\pm1$ & $14.6\pm0.6$\\
$B^+ \rightarrow \mpsiprime \,K^+$ &  $\mpsiprime\rightarrow \mu^+\,\mu^-$ &
          $\phantom{0}71.9\pm13.4$ & $5281\pm3$ & $13.1\pm2.2$\\
$B^+ \rightarrow \mpsiprime \,K^+$ &
	$\mpsiprime\rightarrow\mJpsi\,\pi^+\,\pi^-$ &
          $\phantom{0}37.4\pm\phantom{0}7.4$ & $5279\pm2$ & $11.0\pm2.0$\\
$B^0 \rightarrow \mJpsi \,\mKstarzero$ & $\mJpsi\rightarrow \mu^+\,\mu^-$ &
          $378.8\pm24.8$ & $5280\pm1$ & $13.9\pm1.1$\\
$B^0 \rightarrow \mpsiprime \,\mKstarzero$ &  $\mpsiprime\rightarrow
\mu^+\,\mu^-$ &
          $\phantom{0}20.9\pm\phantom{0}7.3$ & $5275\pm3$ &
		$\phantom{0}8.2\pm2.4$\\
$B^0 \rightarrow \mpsiprime \,\mKstarzero$ & 
$\mpsiprime\rightarrow\mJpsi\,\pi^+\,\pi^-$ &
          $\phantom{0}29.1\pm\phantom{0}7.5$ & $5285\pm4$ & $11.9\pm3.3$\\
\end{tabular}
\end{center}
\label{tab: signal fit params}
\end{table}

\begin{table}
\caption{
A summary of the absolute products of the geometric and kinematic
acceptances, calculated for each decay mode for $B$~mesons produced
with $k_T > 5.0$~GeV/$c$ and $|y_b| < 1.1$.  The
uncertainties given are due to Monte Carlo statistics alone.}
\begin{center}
\begin{tabular}{llc}
\multicolumn{1}{c}{$B$-Meson Decay} & \multicolumn{1}{c}{$c\bar{c}$ Mode} &
			Acceptance ($\times 10^{-3}$) \\ \hline
$B^+\to J/\psi\,K^+$ & $J/\psi\to\mu^+\,\mu^-$ & $19.2\pm 0.3$ \\
$B^+\to\psi(2S)\,K^+$ & $\psi(2S)\to\mu^+\,\mu^-$ & $21.8\pm 0.3$ \\
$B^+\to\psi(2S)\,K^+$ & $\psi(2S)\to J/\psi\,\pi^+\,\pi^-$ &
			$\phantom{0}6.3\pm 0.2$ \\ 
$B^0\to J/\psi\,K^*(892)^0$ & $J/\psi\to\mu^+\,\mu^-$ & 
$\phantom{0}7.8\pm 0.2$ \\
$B^0\to\psi(2S)\,K^*(892)^0$ & $\psi(2S)\to\mu^+\,\mu^-$ &
			$\phantom{0}8.3\pm 0.2$ \\
$B^0\to\psi(2S)\,K^*(892)^0$ & $\psi(2S)\to J/\psi\,\pi^+\,\pi^-$ &
			$\phantom{0}3.9\pm 0.1$ \\
\end{tabular}
\end{center}
\label{tab: acceptances}
\end{table}

\begin{table}
\caption{A summary of the geometric and kinematic acceptance ratios
with associated systematic uncertainties, combined in quadrature, from
the following sources: Monte Carlo statistics, the generated $B$-meson
$P_T$ spectrum, trigger effects, helicity effects, and the CDF
detector simulation uncertainty.}
\label{tab:rel_geom_eff_syst}
\begin{center}
\begin{tabular}{lccc}
~~~~~~~~~~~~~~~~~~Denominator &
\raisebox{-12pt}[0.0in][0.0in]{\parbox{4.0cm}{\begin{eqnarray*}
B^+&\to&J/\psi\,K^+ \\
   &   & \ \mrightdownarrow \,\mu^+\,\mu^-
\end{eqnarray*}}} &
\raisebox{-12pt}[0.0in][0.0in]{\parbox{4.0cm}{\begin{eqnarray*}
B^0&\to&J/\psi\,K^*(892)^0 \\
   &   & \ \mrightdownarrow \,\mu^+\,\mu^-
\end{eqnarray*}}} &
\raisebox{-12pt}[0.0in][0.0in]{\parbox{4.0cm}{\begin{eqnarray*}
B^+&\to&\psi(2S)\,K^+ \\
   &   & \ \mrightdownarrow \,\mu^+\,\mu^-
\end{eqnarray*}}} \\
\unitlength1pt
\begin{picture}(0,0)
\put(0,26){\line(3,-1){135}} 
\end{picture}
\phantom{0} &   &     & \\
Numerator &     &     & \\
\hline
\parbox{1.0cm}{\begin{eqnarray*}
B^0&\to& J/\psi\,K^*(892)^0 \\
   &   & \ \mrightdownarrow \, \mu^+\,\mu^-
\end{eqnarray*}}&
        $0.405\pm 0.025$& & \\

\parbox{1.0cm}{\begin{eqnarray*}
B^+&\to&\psi(2S)\,K^+ \\
   &   & \ \mrightdownarrow \, \mu^+\,\mu^-
\end{eqnarray*}}&
        $1.137\pm 0.084$&
	$2.810\pm 0.285$& \\

\parbox{1.0cm}{\begin{eqnarray*}
B^+&\to& \psi(2S)\,K^+ \\
   &   & \ \mrightdownarrow \, J/\psi\,\pi^+\,\pi^-
\end{eqnarray*}}& 
        $0.329\pm 0.017$&
        $0.813\pm 0.058$&
        $0.289\pm 0.026$\\

\parbox{1.0cm}{\begin{eqnarray*}
B^0&\to&\psi(2S)\,K^*(892)^0 \\
   &   & \ \mrightdownarrow \, \mu^+\,\mu^-
\end{eqnarray*}}&
        $0.431\pm 0.044$&
        $1.066\pm 0.092$&
        $0.379\pm 0.030$\\

\parbox{1.0cm}{\begin{eqnarray*}
B^0&\to&\psi(2S)\,K^*(892)^0 \\
   &   & \ \mrightdownarrow \, J/\psi\,\pi^+\,\pi^-
\end{eqnarray*}}& 
        $0.201\pm 0.017$&
        $0.496\pm 0.032$&
        $0.177\pm 0.023$
\end{tabular}
\end{center}
\end{table}

\begin{table}
\caption{A summary of the efficiency-product ratios and their
associated systematic uncertainties from the following sources: track
reconstruction efficiencies, confidence-level criteria
efficiencies, and decay-length requirement efficiencies.}
\label{tab:rel_all_eff_syst}
\begin{center}
\begin{tabular}{lccc}
~~~~~~~~~~~~~~~~~~Denominator &
\raisebox{-12pt}[0.0in][0.0in]{\parbox{4.0cm}{\begin{eqnarray*}
B^+&\to&J/\psi\,K^+ \\
   &   & \ \mrightdownarrow \,\mu^+\,\mu^-
\end{eqnarray*}}} &
\raisebox{-12pt}[0.0in][0.0in]{\parbox{4.0cm}{\begin{eqnarray*}
B^0&\to&J/\psi\,K^*(892)^0 \\
   &   & \ \mrightdownarrow \,\mu^+\,\mu^-
\end{eqnarray*}}} &
\raisebox{-12pt}[0.0in][0.0in]{\parbox{4.0cm}{\begin{eqnarray*}
B^+&\to&\psi(2S)\,K^+ \\
   &   & \ \mrightdownarrow \,\mu^+\,\mu^-
\end{eqnarray*}}} \\
\unitlength1pt
\begin{picture}(0,0)
\put(0,26){\line(3,-1){135}} 
\end{picture}
\phantom{0} &   &     & \\
Numerator &     &     & \\
\hline
\parbox{1.0cm}{\begin{eqnarray*}
B^0&\to& J/\psi\,K^*(892)^0 \\
   &   & \ \mrightdownarrow \, \mu^+\,\mu^-
\end{eqnarray*}}&
        $0.930\pm 0.030$& & \\

\parbox{1.0cm}{\begin{eqnarray*}
B^+&\to&\psi(2S)\,K^+ \\
   &   & \ \mrightdownarrow \, \mu^+\,\mu^-
\end{eqnarray*}}&
        $0.980\pm 0.023$&
	$1.05\pm 0.03$& \\

\parbox{1.0cm}{\begin{eqnarray*}
B^+&\to& \psi(2S)\,K^+ \\
   &   & \ \mrightdownarrow \, J/\psi\,\pi^+\,\pi^-
\end{eqnarray*}}& 
        $0.703\pm 0.055$&
        $0.755\pm 0.062$&
        $0.717\pm 0.056$\\

\parbox{1.0cm}{\begin{eqnarray*}
B^0&\to&\psi(2S)\,K^*(892)^0 \\
   &   & \ \mrightdownarrow \, \mu^+\,\mu^-
\end{eqnarray*}}&
        $0.923\pm 0.030$&
        $0.992\pm 0.024$&
        $0.942\pm 0.031$\\

\parbox{1.0cm}{\begin{eqnarray*}
B^0&\to&\psi(2S)\,K^*(892)^0 \\
   &   & \ \mrightdownarrow \, J/\psi\,\pi^+\,\pi^-
\end{eqnarray*}}& 
        $0.651\pm 0.053$&
        $0.700\pm 0.055$&
        $0.664\pm 0.054$
\end{tabular}
\end{center}
\end{table}

\begin{table}
\caption{ Branching fractions of the daughter-meson decay modes
reconstructed in the present analysis.  The world-average branching
fractions were used for the charmonium mesons~\protect\cite{ref: PDG}.
Note that the branching fraction for the
$\mpsiprime\rightarrow\mu^+\,\mu^-$\ decay assumes lepton
universality.  The \Kstarzero\ branching fraction is based on isospin
symmetry.}
\label{tab: secondary BR}
\begin{center}
\begin{tabular}{lc}        
\multicolumn{1}{c}{Decay Mode} &
		\multicolumn{1}{c}{Branching Fraction} \\ \hline
$J/\psi\to\mu^+\,\mu^-$   & $(6.01 \pm 0.19) \times 10^{-2}$ \\
$\psi(2S)\to\mu^+\,\mu^-$ & $\phantom{00}(8.5  \pm 0.7)  \times 10^{-3}$ \\
$\psi(2S)\to J/\psi\,\pi^+\,\pi^-$ &
		$(3.07 \pm 0.19) \times 10^{-1}$ \\ 
$K^*(892)^0\to K^+\,\pi^-$ & $2/3$ \\
\end{tabular}
\end{center}
\end{table}

\begin{table}
\caption{The measured ratios of branching fractions times fragmentation
fractions for the four decay modes.  The systematic uncertainties have
been calculated taking into account cancellations and correlations in
uncertainties.}
\label{tab: final results}
\begin{center}
\begin{tabular}{cc}
Quantity & Measured Ratio \\ \hline
${{f_d}\over{f_u}}\,\cdot\,
{{
\mBR(B^0 \rightarrow J/\psi \, \mKstarzero)
               }\over{
\mBR(B^+\rightarrow J/\psi \, K^{+})
						}}$
&   $1.76\pm0.14\pm0.15$ \\[0.10in]
${{
 \mBR(B^+ \rightarrow \mpsiprime \,K^{+})
               }\over{
 \mBR(B^+\rightarrow \mJpsi \,K^{+})
						}}$
& $ 0.558\pm0.082\pm0.056$  \\[0.10in]
${{f_d}\over{f_u}}\,\cdot\,
{{
\mBR(B^0 \rightarrow \mpsiprime \,\mKstarzero)
               }\over{
\mBR(B^+\rightarrow J/\psi \,K^{+})
						}}$
&   $0.908\pm0.194\pm0.100$ \\[0.10in]
${{f_u}\over{f_d}}\,\cdot\,
{{
\mBR(B^+ \rightarrow \mpsiprime\,K^+)
               }\over{
\mBR(B^0 \rightarrow \mJpsi\,\mKstarzero)
						}}$
&    $0.317\pm0.049\pm0.036$ \\[0.10in]
${{
\mBR(B^0 \rightarrow \mpsiprime\,\mKstarzero)
               }\over{
\mBR(B^0 \rightarrow \mJpsi\,\mKstarzero)
						}}$
&    $0.515\pm0.113\pm0.052$ \\[0.10in]
${{f_d}\over{f_u}}\,\cdot\,
{{
\mBR(B^0 \rightarrow \mpsiprime\,\mKstarzero)
               }\over{
\mBR(B^+ \rightarrow \mpsiprime \,K^+)
						}}$
&    $1.62\pm0.41\pm0.19$
\end{tabular}
\end{center}
\end{table}

\end{document}